\documentclass[lettersize,journal]{IEEEtran}
\usepackage{amsmath,amsfonts}
\usepackage{algorithmic}
\usepackage{algorithm}
\usepackage{array}
\usepackage[caption=false,font=normalsize,labelfont=sf,textfont=sf]{subfig}
\usepackage{textcomp}
\usepackage{stfloats}
\usepackage{url}
\usepackage{verbatim}
\usepackage{graphicx}
\usepackage{cite}
\usepackage{siunitx}
\usepackage{amssymb}

\begin{document}

\title{Voluminous Fur Stroking Experience through Interactive Visuo-Haptic Model in Virtual Reality}

\author{Juro Hosoi, Du Jin, Yuki Ban, and Shin'ichi Warisawa

        % <-this % stops a space

\thanks{Manuscript received X X, X; revised X X, X.This work was supported by JSPS KAKENHI Grant Numbers JP23H04333, JP21H03478.}
\thanks{Juro Hosoi is with the Graduate School of Frontier Sciences, The University of Tokyo, Chiba 277-8563, Japan (e-mail: hosoijuro@lelab.t.u-tokyo.ac.jp).}
\thanks{Du Jin is with the Graduate School of Frontier Sciences, The University of Tokyo, Chiba 277-8563, Japan (e-mail: dujin@lelab.t.u-tokyo.ac.jp).}
\thanks{Yuki Ban is with the Graduate School of Frontier Sciences, The University of Tokyo, Chiba 277-8563, Japan (e-mail: ban@edu.k.u-tokyo.ac.jp).}
\thanks{Shin'ichi Warisawa is with the Graduate School of Frontier Sciences, The University of Tokyo, Chiba 277-8563, Japan (e-mail: warisawa@edu.k.u-tokyo.ac.jp).}% <-this % stops a space
% https://kaken.nii.ac.jp/ja/grant/KAKENHI-PUBLICLY-23H04333/
% https://kaken.nii.ac.jp/ja/grant/KAKENHI-PROJECT-21H03478/
\thanks{This work has been submitted to the IEEE for possible publication. Copyright may be transferred without notice, after which this version may no longer be accessible.}}

\markboth{Journal of \LaTeX\ Class Files,~Vol.~X, No.~X, XXX~YYY}%
{Juro Hosoi \MakeLowercase{\textit{et al.}}: Voluminous Fur Stroking Experience through Interactive Visuo-Haptic Model in Virtual Reality}


\maketitle

\begin{abstract}
% tran 150-250 words
% 非専門家でも理解できるような、一般向けのバックグラウンド。1 - 2 文。Background をやや狭い範囲に絞る文章、2 - 3 文。Problem の特定。シンプルに 1 文で。この論文が、その problem に対してどのように答えを出しているか。結果を述べる部分。結果の解釈、1 - 2 文。2 - 3 文で、全体を broader perspective で説明するような文章。
The tactile sensation of stroking soft fur, known for its comfort and emotional benefits, has numerous applications in virtual reality, animal-assisted therapy, and household products.  
Previous studies have primarily utilized actual fur to present a voluminous fur experience that poses challenges concerning versatility and flexibility.
In this study, we develop a system that integrates a head-mounted display with an ultrasound haptic display to provide visual and haptic feedback. Measurements taken using an artificial skin sheet reveal directional differences in tactile and visual responses to voluminous fur. 
Based on observations and measurements, we propose interactive models that dynamically adjust to hand movements, simulating fur-stroking sensations.
Our experiments demonstrate that the proposed model using visual and haptic modalities significantly enhances the realism of a fur-stroking experience.
Our findings suggest that the interactive visuo-haptic model offers a promising fur-stroking experience in virtual reality, potentially enhancing the user experience in therapeutic, entertainment, and retail applications.
\end{abstract}

\begin{IEEEkeywords}
Fur, Hair, Haptic display, mid-air haptics, Virtual Reality
\end{IEEEkeywords}

\section{Introduction}

\IEEEPARstart{H}{aptics} plays a crucial role in human cultural and social activities. Studies on haptics consider various perspectives: perception, sensing, and feedback. 
 Particularly, the touch of fur, such as that of animals, is comfortable and there is a demand for stroking and touching fur in our daily lives~\cite{Ujitoko2022, Ujitoko2023}.
 The experience of touching fur or animals has been utilized in various applications including robots and furniture to evoke a sense of closeness and comfort~\cite{Ueki2007, Wada2005, Yohanan2012, Geva2020}; entertainment; animal-assisted therapy~\cite{Martin2002, Nimer2007, Martin2009, Kamioka2014}; treatment for overcoming animal phobias in virtual reality (VR) applications~\cite{Alvear2017}; and online shopping for clothing, towels, and interiors~\cite{Alka2003, Ornati2020, Kitagishi2023}.

%\red{\begin{CJK}{UTF8}{ipxm}[先行研究の課題] 実際の毛を用いるのは大変．疑似提示はフラットなものばっか（´・ω・｀）\end{CJK}}
%\cyan{\begin{CJK}{UTF8}{ipxm}どんな取り組みが行われてきて，その上で未だ達成できてない課題は？    その課題の重要性・影響力・本質的なムズポイントは？(自分で無理やり課題を作り出してない？ やるだけ笑なのでやられてないだけじゃない？)\end{CJK}}

%\begin{CJK}{UTF8}{ipxm}毛並みを触る体験を創出するため，様々な研究がこれまでに進められてきたが，先行研究で取られてきたアプローチは大きく2つに分けることができる．一つ目は実際の毛を用いる手法である．Wadaらは毛でもふもふのぬいぐるみを導入することで，老人の気分を長期間にわたって改善した．またLeeらは毛束を把持する機構を工夫することで毛並みの角度や高さといった毛並みのプロパティの一部を変化させた．これらの手法は実際の毛並みを用いているため，高いリアリティを提示できる反面，提示したい毛並みと準備する毛並みのプロパティ(例えば，密度や細さ)をある程度揃える必要がある上，提示するコンテンツに合わせて位置や向き・形状を逐一そろえる必要があり，汎用性や柔軟性に欠けてしまう．それに対する２つ目の手法は，疑似的に毛並み触感を代替的に提示する手法である．疑似的に毛並みとの接触を提示することで，物理的な制約を飛ばして，様々な応用につなぐことができる．Linらは猫のCGオブジェクトの表面に触れた際に電気刺激を提示する試みを行った．この研究は猫の平面テクスチャへの接触を模擬することを達成している．一方で，我々の知る限り，ボリュームのある柔らかい毛並みの触感提示については先行研究では達成されていない．しかし，猫や犬などの毛並みを触るとき，我々は毛並みが柔らかいことを期待するなど~\cite{Ujitoko2023}，毛並みとのインタラクションを考えた時に毛並みの持つボリューム感は重要であると言える．\end{CJK}

Various studies have been conducted to create the experience of touching fur. The methods used can be broadly divided into two approaches. The first involves the use of real fur. Wada et al. introduced three furry animal robots in a health service facility to improve the mood and depression of elderly people~\cite{Wada2005}. 
Lee et al. developed a mechanism for grasping brush hairs, allowing manipulation of fur characteristics such as angle and height~\cite{Lee2021}. 
These methods employing actual fur offer high levels of realism. However, the specific properties of prepared fur, such as density and fineness, should match the intended display. Further, existing methods lack versatility because they require precise alignment of position, orientation, and shape to match the intended content.
The second approach involves simulating the tactile experience of fur without using actual fur. 
Physical constraints can be reduced by simulating contact with fur, thereby facilitating various applications. Lin et al. conducted the studies on electrical stimulation by touching the surface of a computer graphics (CG) cat~\cite{Weikang2022}. In their research, the simulation of contact with a flat-textured cat surface can be achieved. However, to the best of our knowledge, prior research has not successfully presented the tactile sensation of voluminous and soft fur. When touching the fur of animals including cats and dogs, we expect their fur to feel soft~\cite{Ujitoko2023}, and the sense of volume is considered crucial in interactions with fur.

%\red{\begin{CJK}{UTF8}{ipxm}[本研究の目的] 疑似的なボリュームのある毛並み提示\end{CJK}}
%\begin{CJK}{UTF8}{ipxm}本研究の目的は，疑似的にボリュームのある毛並み触感を提示することである．ボリュームのある毛は単なる2Dテクスチャではなく，形状や向きが存在する上，手の動きによって異なる変形を見せ，異なる触り心地となる．我々はこれらの特徴を整理し，インタラクティブなモデルを構築することで，ボリューミーな毛並みを撫でる体験を疑似的に提示することを提案した．\end{CJK}
This study aims to simulate the tactile experience of stroking voluminous fur. Voluminous fur differs from flat textures because it possesses shape and direction, and exhibits different deformations and tactile sensations based on hand movements and fur conditions, resulting in varied tactile sensations. 
Based on these features, we propose an interactive visuo-haptic model to present a fur-stroking experience.

%\begin{CJK}{UTF8}{ipxm}ボリューミーな毛を実際の毛を用いずに提示することが困難であると考えられる理由は以下の2点である．まず一本一本の毛は非常に細かいため，これをピンアレイデバイスなどで物理的に再現しようとすると，実質的に毛が必要となる．さらに毛が集合したボリューミーな全体は，手の皮膚よりも柔らかく，微小な触感となるため，振動子等のリジットな触覚デバイスの接触が繊細な毛の触感を損なってしまう．これに対し，我々は空中超音波振動子を用いることで非接触に触覚刺激を提示するアプローチを選択した．また，微細な空間マッピングについては，微細なCGモデルを用いることでクロスモーダル錯覚を用いることで解決することを試みた．\end{CJK}

Challenges associated with replicating the tactile experience of voluminous fur without real fur are primarily attributed to two factors.
First, rendering spatial simulations using devices such as pin arrays~\cite{Leithinger2010} is nearly impossible without actual fur. Second, voluminous fur is much softer than human skin and is registered as a subtle tactile sensation. Therefore, rigid haptic devices such as vibrators~\cite{Christoph2004, Choi2013} compromise the delicate feeling of fur. Thus, we use mid-air ultrasound haptic feedback to provide tactile stimuli without contact. 
Furthermore, we use cross-modal illusions with fine CG hair models for precise spatial mapping.

%\begin{CJK}{UTF8}{ipxm}まず，我々は，ボリュームのある毛並みの特徴を観察するため，撫でる体験を模した人工皮膚による測定系を作成した．測定の結果，我々は，毛並みの向きに応じて異なる視覚的・触覚的挙動の異方性を発見した．\end{CJK}
%\begin{CJK}{UTF8}{ipxm}測定を基に，手の動きに対する毛並みの特徴を提示する触覚デザインとCGモデルを作成し，これらを非接触に触覚刺激を提示できる超音波振動子アレイとHMDを用いて提示するシステムを構築した．\end{CJK}
First, we develop a measurement system using an artificial skin sheet to mimic the experience of stroking, thereby enabling the observation of voluminous fur characteristics. Our measurements reveal that visual and tactile responses exhibit anisotropy depending on the stroke direction. Using measurement results, we develop tactile and CG models that present fur texture characteristics in response to hand movements, and construct a system to present these models using an ultrasound haptic display array and a head-mounted display (HMD).

%\red{\begin{CJK}{UTF8}{ipxm}[RQ] インタラクティブな刺激設計が毛並みのリアリティには大事？ \end{CJK}}

This study verifies whether an interactive model using combined visual and haptic feedback related to hand movements can replicate the experience of stroking voluminous fur. In Experiment 1, we assess the effect of our tactile design on evoking fur-stroking sensation, focusing solely on haptic stimuli to eliminate potential anticipatory biases. 
Experiment 2 explores the closeness of our proposed interactive visual and haptic models in reproducing fur-stroking sensation compared to actual fur.

This paper is a revised version of the abstract demonstrated at SIGGRAPH Asia 2023~\cite{Hosoi2023Fur}. 
We have incorporated more detailed measurements and models. 
We conducted two additional experiments to examine the effectiveness of the proposed method.

\section{Related work}

\subsection{Fur Display}

%\begin{CJK}{UTF8}{ipxm}毛とのインタラクションは，文化的・社会的に意味からも非常に重要である．\end{CJK}
%\begin{CJK}{UTF8}{ipxm}モチベ・活用先も無限にある．\end{CJK}

%\begin{CJK}{UTF8}{ipxm}提示に関して，視覚・触覚的にいろいろやられているデカ分野．ボリュームのある触感はまだ．\end{CJK}

%\begin{CJK}{UTF8}{ipxm}本研究は，ボリュームのある毛を疑似的に出した初めての研究である．あと測定とか初期モデルを作ったのも貢献でかいよ．\end{CJK}

The tactile experience of touching fur has profound significance across various domains. For example, therapeutic robots with fur have been used in health service facilities to improve the mood of elderly people \cite{Wada2005}. Fur-covered furniture at home has been demonstrated to increase pleasantness and comfort \cite{Ueki2007}. Furry animals, such as dogs and cats, have been used extensively in animal-assisted therapy to improve emotional well-being \cite{Nimer2007}.

Previous studies on fur displays predominantly used real fur. Furukawa et al. designed a fur interface with a bristling effect using a piece of real fur equipped with vibrational motors \cite{Furukawa2009, Furukawa2010}. Lee et al. proposed a handheld fur display that presented different stiffness, roughness, and surface height by controlling fur length and bending fur direction \cite{Lee2021}.
Nakajima et al. integrated plastic fiber optics into a fur display to provide visual feedback in addition to tactile sensations during interactions \cite{Nakajima2011}. Although the use of real fur can provide the most realistic tactile sensation, systems become less adaptable because the fur properties are restricted to those of the fur used. In addition, if such a type of fur display is used to provide haptic feedback of a virtual furry object in VR, the exact positioning, orientation, and shaping of real fur require adjustment in real time to correspond with the intended effect. Generally, such a feature is not feasible for several interactive applications such as petting a moving virtual animal.

A study presented fur tactile sensation without the use of real fur or fur-like devices. Instead, a wearable electro-tactile rendering system was designed to simulate the tactile sensation of stroking a furry cat using electrical stimulators on a haptic glove \cite{Weikang2022}. However, fur was treated as a flat-textured surface, and the complex properties of fur and its interaction with the hands of the user in 3D space were not considered.

Our proposed method fills the research gap by presenting voluminous fur tactile sensations without the use of real fur while considering complex fur properties and 3D interactions based on a kinematic model elicited from real fur.

\subsection{Ultrasound Haptic Feedback}

%\red{\begin{CJK}{UTF8}{ipxm}超音波でいろいろ出せそうなことは知られている\end{CJK}}

%\begin{CJK}{UTF8}{ipxm}超音波は低遅延で非接触で位置の指定が用意．air bubble vibrationにより，強度，摩擦感，圧力などの提示が行われてきた．\end{CJK}

%\begin{CJK}{UTF8}{ipxm}モデルを組んで超音波出力を出すも色々やられている．力のシミュはばねモデル．位置のシミュは水とか柔らかいゴム．\end{CJK}

Ultrasound haptic devices have emerged as promising solutions for providing touch sensations without direct physical contact. These devices use focused sound waves to provide multi-point stimulation across a large area with high temporal and positional accuracy, enabling unique tactile experiences that are not possible with other technologies \cite{rakkolainen2020survey}.

Since the introduction and evaluation of ultrasound haptic in 2008 \cite{iwamoto2008non}, it has been extensively used in virtual user interface components to enhance gesture input interactions. For example, Sand et al. installed an ultrasonic phased array on an HMD to provide tactile feedback when pressing a button on a virtual keyboard \cite{sand2015head}. The participants in their user study unanimously found the feedback preferable. Harrington et al. used ultrasound haptic feedback for a slider bar in a driving simulator, which resulted in the shortest interaction time and the highest number of correct responses \cite{harrington2018exploring}. For such types of interactions, the existence of haptic feedback is more important than how it feels, as the main purpose is to confirm to users that the system is reacting to their input \cite{rakkolainen2020survey}. Thus, ultrasound haptic feedback is a popular choice owing to its flexibility and non-contact nature.

Recently, the technology has been used to simulate complex haptic sensations in the real world. Jang et al. developed a fluid tactile rendering method using an ultrasound haptic display and smoothed-particle hydrodynamics to provide realistic vibro-tactile feedback for virtual fluid interactions on the hands of users \cite{Jang2020}. Singhal et al. integrated heat modules with an ultrasound haptic display in an open-top chamber to generate thermos-tactile feedback resembling steam or campfire \cite{Singhal2023}. Motoyama et al. presented a non-contact cooling sensation using ultrasound-driven mist vaporization \cite{Motoyama2022}.
Such studies suggest that the capacity of ultrasound haptic feedback can be further expanded using appropriate modulation methods and complementary modalities.

Our study utilizes ultrasound haptic feedback to present the tactile sensation of voluminous fur, which is complex and remains unexplored. We anticipate this novel application to broaden the scope and applicability of ultrasound technology in immersive environments.

\section{Approach}

To design a visuo-haptic model for stroking voluminous fur, measuring the visual characteristics and reactive forces involved is crucial. Although previous research has focused on measuring a single hair or short fur~\cite{Watzky2011, Camillieri2012, BUENO2013, Michel2015}, sufficient measurements have not been conducted for voluminous fur. Therefore, we conducted measurements on behavior while stroking voluminous fur on artificial skin for this study. 
Subsequently, leveraging these observations, we constructed visual and haptic models of voluminous fur.

\subsection{Measurement System and Procedure}
\label{subsec_measurment}

\begin{figure}[!htbp]
    \centering
    \includegraphics[width=\linewidth]{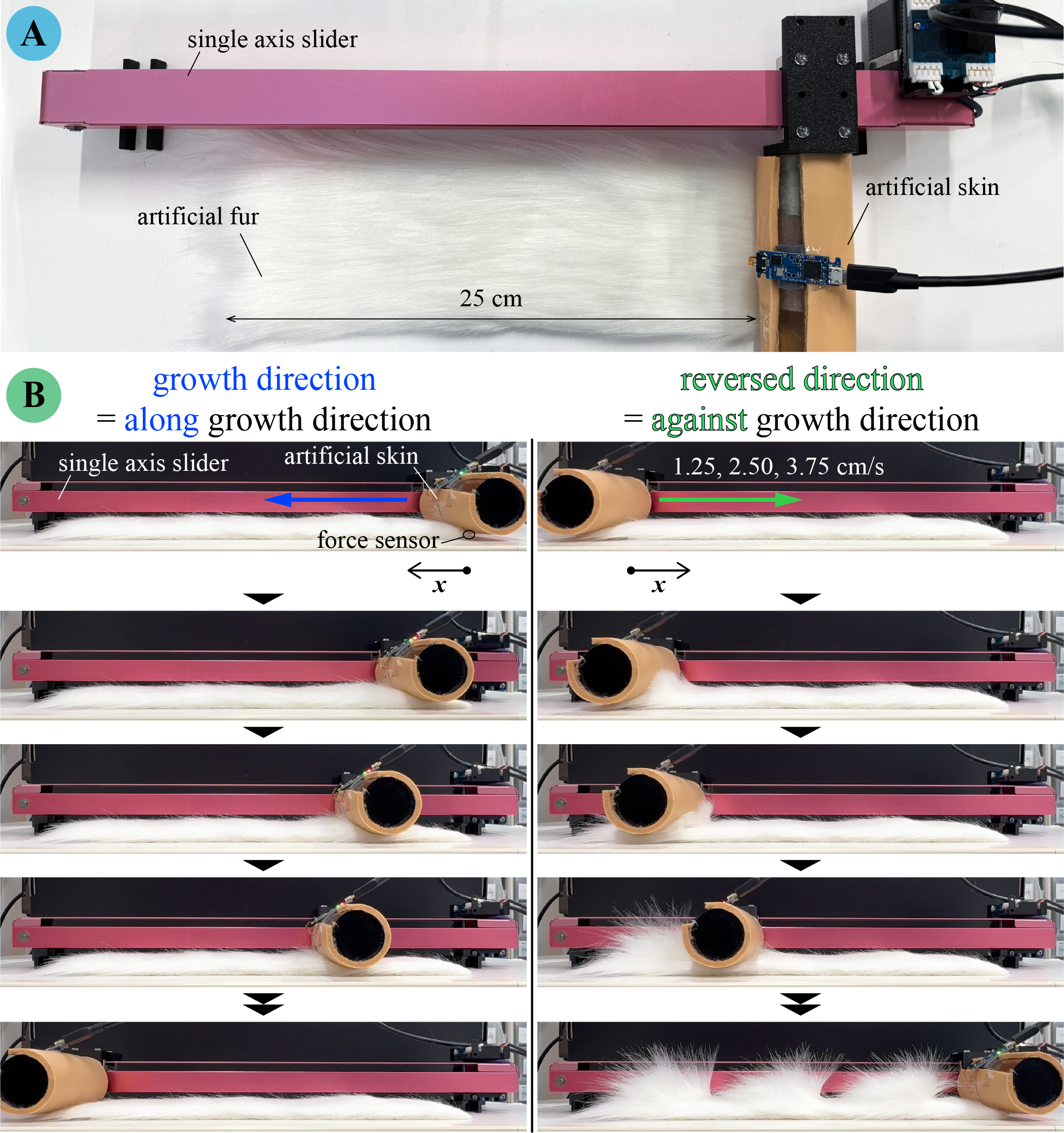}
    \caption{(A) Top view of the measurement system for observing visual behavior and resistive force when stroking fur of width \SI{25}{cm}. (B) Side view of the measurement system stroking voluminous fur. Two vertical rows of pictures show the transition of visual behavior when the fur is stroked along and against the growth direction, respectively.}
    \label{fig_sens_stroke}
\end{figure}

%\begin{CJK}{UTF8}{ipxm}ボリュメトリックな毛を撫でた際の視覚・触覚的な挙動を観察するため，髪の毛をなぞった際の力を測定するセンサ系を参考に，測定系を構築した．\end{CJK}
To investigate visual and haptic behaviors during voluminous fur-stroking, we developed a measurement system by referencing a sensor system that measures the force of hair strands~\cite{Okuyama2011, Kakizawa2013, Kakizawa2016}.
Figure~\ref{fig_sens_stroke} (A) shows the top view of the measurement system.
The measurement system comprised a cylindrical resin (diameter: \SI{30}{mm}) equipped with an artificial skin sheet (Shore A hardness: 10--15 mm; thickness: \SI{4}{mm}), a force sensor (calibrated \SI{8}{mm} diameter sensor \SI{1}{\newton}/\SI{0.22}{lbf}, SingleTact), single-axis motor for movement (PCS9S-330-S330, THK, stroke length: \SI{330}{mm}), and artificial fur (hair length: \SI{5}{cm}; fur width: \SI{25}{cm}).

We positioned the force sensor \SI{1}{cm} above the lower edge of the fur and moved it in two horizontal directions, growth and reversed directions, at \SI{1.25}{cm/s}, \SI{2.50}{cm/s}, and \SI{3.75}{cm/s}. 
We recorded the visual response using a camera and measured the vertical force using the force sensor. 
Before each movement, we ensured uniformity of the fur texture by combing it.

\subsection{Observations of Visual Behavior}
\label{subsec_measument_visual}

First, we discuss the visual characteristics of voluminous fur based on our observations. Figure~\ref{fig_sens_stroke} (B) illustrates the temporal changes in fur when stroked in two directions. The left column represents the behavior when stroked along the natural direction of growth, whereas the right column represents the behavior when stroked against the growth direction of fur. As shown in the figure, the visual behavior of fur varied during and after the stroke, depending on the stroke direction. However, little difference in visual behavior was observed depending on stroke speed.

When using a slider to stroke hair strands, it is known that the behavior of the hair can be divided into two primary phases: the deformation phase and the rubbing phase~\cite{Michel2015}. 
During the deformation phase, the slider lays down and bends the hair. In the rubbing phase, the slider rubs upon the bent hair. In this study, for fur with a high density of hairs, this behavior is expected to occur continuously.

When stroked along the growth direction, the fur strands were compressed by the moving artificial skin. Subsequently, the fur strands regained their original shape after the skin passed over them. Consequently, minimal visible traces remained after the stroking motion. 
Conversely, when voluminous fur was stroked against its growth direction, the fur bundles that were initially in contact with the hand were gradually lifted to the opposite side and they started to curl, involving other fur bundles along the stroke path. As the experiment progressed, the bent fur bundles were released and the skin made contact with the next bundle that was lifted. The released bundles maintained their curled shape, and the repetitive process left cyclic traces of stroking. The unique behavior of maintaining a curled shape has been noted in previous research for potential applications in visual displays~\cite{Sugiura2014, Horishita2014, Sugiura2023}.

\subsection{Measurement Results of Resistive Force}
\label{subsec_measument_haptic}

%\red{\begin{CJK}{UTF8}{ipxm}[測定結果] \end{CJK}}

%\begin{CJK}{UTF8}{ipxm}視覚的挙動の観察に合わせて計測した力について議論する．図1は撫でる方向と速度を変えて計測した垂直方向の反力の結果である．横軸のxはセンサが毛並みに触れた点を原点とした時の，撫でた位置を意味する．視覚と同様に触覚方面でも異方性が見られた．一方で，速度による影響はほぼ見えなかった．順目方向に撫でた時は，定常な力が見られた．一方で，逆目方向で撫でた際は，視覚的挙動と同様にサイクリックな挙動が見えた．一つのサイクルに着目すると，反力は最初の変形フェーズに急激に大きくなり，その後rubbingパートでゆるやかになる．この挙動は直立した数個のpileの表面を計測した先行研究との結果とも一致している．先行研究と異なる点として，最初から倒れた状態から裏返りまでやっている点毛が長くまた密度が高いためサイクリックな挙動が見られる点である．サイクルは8cm周期で起きており，これは5cmの毛が倒れた状態から逆側に反り返った状態までの距離にあたる\end{CJK}

\begin{figure}[!htbp]
    \centering
    \includegraphics[width=\linewidth]{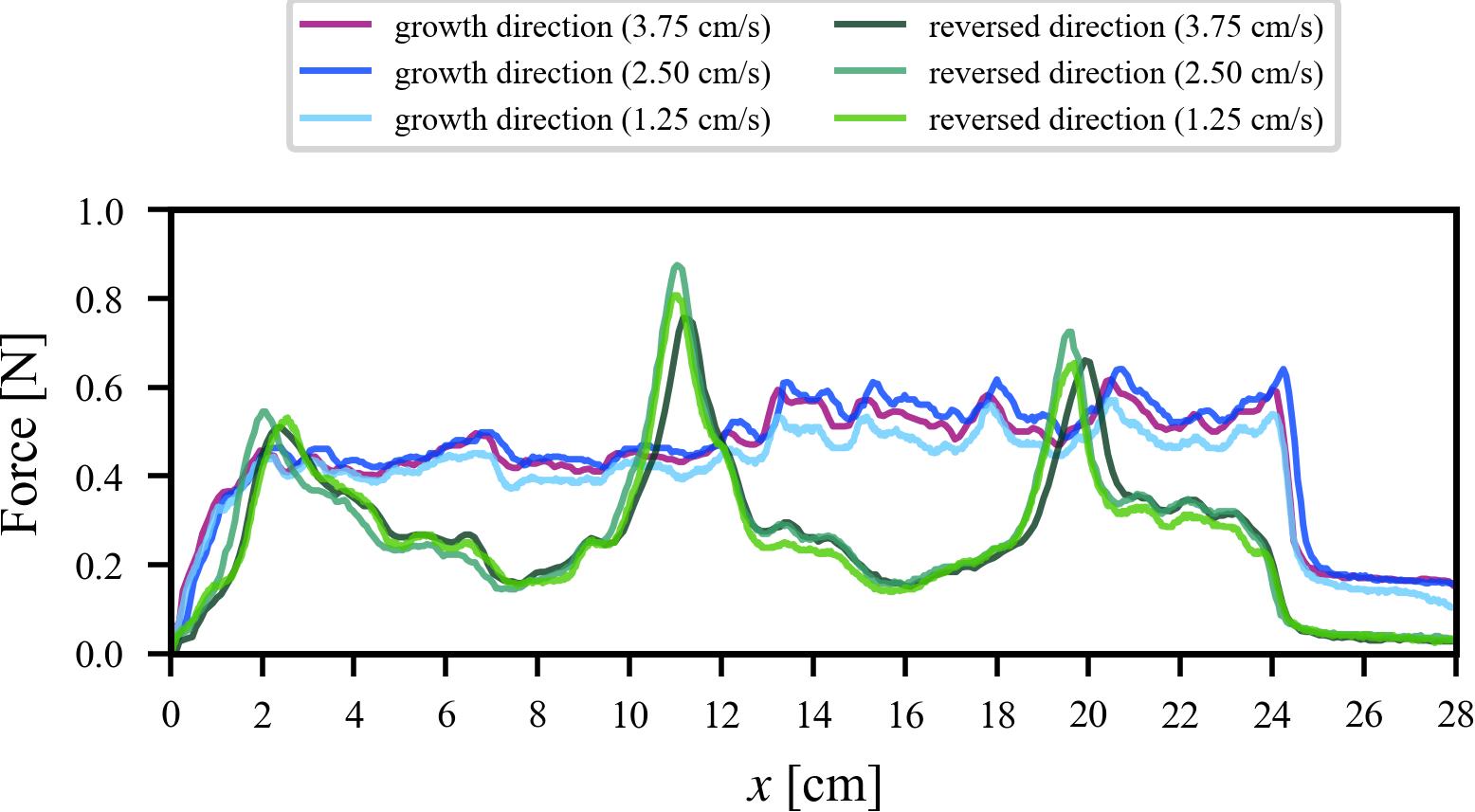}
    \caption{Results of the measured vertical force when the force sensor comes in contact with the fur at the point designated as $x = 0$. The horizontal width of fur is \SI{25}{cm}.
    Each line in the graph represents six distinct conditions, a combination of three velocity conditions and two movement direction conditions. }
    \label{fig_sens_result}
\end{figure}

We examined the forces measured alongside visual behavioral observations. Figure~\ref{fig_sens_result} shows the results of the vertical forces measured in multiple stroke directions and speeds. The x-axis represents the stroke position relative to the point where the sensor first touches the fur. Anisotropy was observed in the haptic aspect, similar to the visual behavior. Small effects due to speed were also observed. When fur was stroked along the growth direction, steady vertical forces were observed. Conversely, when stroked in the reversed direction, a cyclic behavior similar to the visual behavior was observed. In a single cycle, forces sharply increased during the deformation phase and then gradually decreased during the rubbing phase. The behavior aligned with the results of previous studies that measured the surfaces of several upright piles~\cite{Michel2015}. Notably, our research highlights the cyclic behavior owing to long and dense fur transitioning from a flat to a flipped state. 
Each cycle occurred approximately every \SI{8}{cm}, corresponding to the distance for a bundle of hairs, \SI{5}{cm} in length and approximately \SI{3}{cm} in width, collapsing and flipping.

\subsection{Visual Model of Voluminous Fur}
\label{subsec_visual_model}

%\begin{CJK}{UTF8}{ipxm}毛並みの視覚モデルはCG分野で数多く提案されている．毛並みのモデルは複雑かつ量が多く，計算コストも大きくなりがちなので，インタラクティブに動くわけでなく単なる静的なテクスチャとして実装されている場合も多い．インタラクティブなモデルは，基本的にはコリジョンが発生するオブジェクトにつられて動き，リリースされたのちに徐々に初期の角度に戻るような設計がされている．本研究では，3.3節のボリューミーな毛の視覚的挙動の観察結果に基づき，リリースされた点に応じて収束する形状を変えることで，逆立つ現象を実装した．図15は，実装した毛を2方向から手で撫でている際の映像である．順目で撫でている際は下に潰れ緩やかに戻るのに対し，逆目方向で撫でている際は，逆立つ挙動を見せている．\end{CJK}

\begin{figure}[!htbp]
    \centering
    \includegraphics[width=\linewidth]{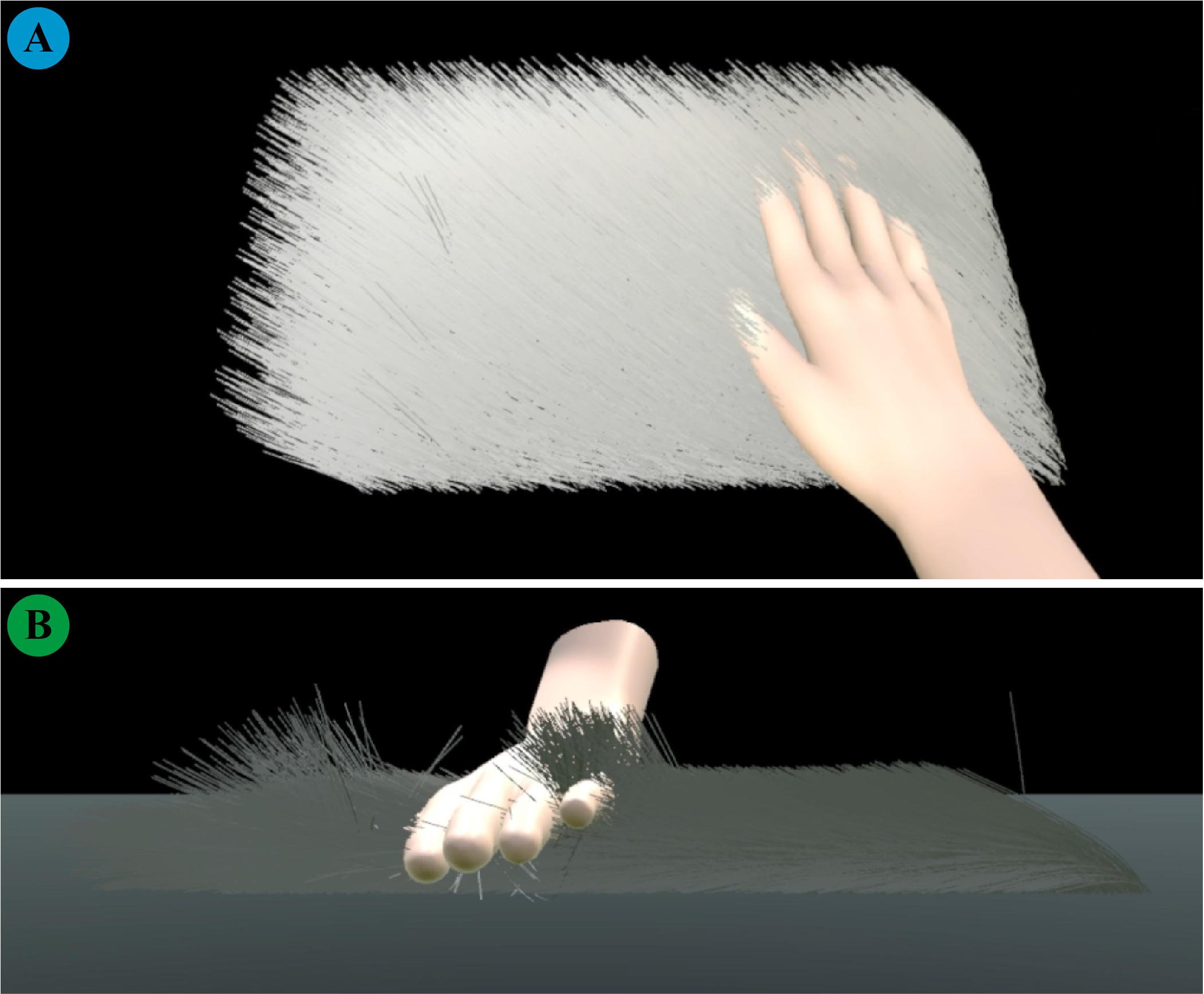}
    \caption{(A) Top view of the implemented fur CG model stroked in the growth direction. The fur is pressed down by the hand.
    (B) Side view of the fur CG model stroked against the growth direction. The fur is lifted by the hand and it stands up after the hand passes.}
    \label{fig_visual_model}
\end{figure}

Various fur visual models have been proposed using CG~\cite{Petrovic2005, McAdams2009, Muller2012, Chai2014, Jiang2017}. Owing to their complexity and high computational costs, fur models are often implemented as static textures in common applications.
Interactive models typically move in response to colliding objects and are designed to gradually return to their initial angles after release. 
In this study, we used the Unity hair simulation system~\cite{UnityHair} to implement the phenomenon of hair standing at the end by adjusting the shape to converge based on a release point, according to the observation of visual behavior of voluminous fur in \ref{subsec_measument_visual}. 
Figure~\ref{fig_visual_model} shows the footage of manually stroking the implemented fur from two directions. When stroked forward, the fur collapsed downward and gradually returned, whereas stroking backward resulted in it standing at the end.

%\begin{CJK}{UTF8}{ipxm}本研究はCG研究ではないよ\end{CJK}

Our CG model achieved a frame rate of 72 frames per second on HMD (Meta Quest 3, Meta) using a laptop PC equipped with NVIDIA GeForce RTX 3080 Laptop GPU, 16 GB of system RAM, and an Intel Core i7-11800H processor.

\subsection{Haptic Model of Voluminous Fur}
\label{subsec_haptic_model}

%https://www.ultraleap.com/datasheets/STRATOS_Inspire_Data_Sheet_April.pdf
To present the soft tactile sensation of voluminous fur, we selected a mid-air ultrasound tactile display capable of offering non-contact tactile stimuli, rather than rigid haptic devices. 
For this study, we used Ultrahaptics STRATOS Inspire equipped with 256 ultrasonic transducers and a Leap Motion Controller for hand tracking. The intensity of the mid-air ultrasound haptic feedback was adjustable from 0.0 to 1.0, with 1.0 corresponding to \SI{10}{mN} on a \SI{2.1}{cm} circular target. 
Spatiotemporal modulation (STM)~\cite{Frier2019, Hasegawa2022, Mendes2024} was employed to the focus point, moving it in a circle of circumference \SI{20}{cm}, a common size in previous research on ultrasound haptic feedback~\cite{Ablart2019, Wojna2023, Wojna2023_par, Shen2023} and sufficiently small compared to the palm size of an adult male hand~\cite{Komandur2009}. Regarding STM, the primary parameters comprise intensity and frequency, aside from the geometric shape. Herein, intensity refers to the magnitude of the applied force, and frequency refers to the speed at which the focal point moves along a circle.

%\begin{CJK}{UTF8}{ipxm}まず，3節の計測結果を基に超音波触覚刺激の強度設計について議論する．順目方向に毛を撫でている際は，垂直反力は一定であるため，図2から0.6とした．一方，逆目方向で毛を撫でている時，反力はサイクリックな挙動を見せる．この挙動を図解したものが図3である．皮膚は毛のbundleを持ち上げて，逆立て，リリースする．毛の長さを$l$, 毛のbundleの幅を$b$とすると，撫でる前の初期状態はほぼ水平なので 垂直反力$F$は$l+b$の周期関数と考えることができる.\ref{eq_macromodel}\end{CJK}

\begin{figure}[!t]
    \centering
    \includegraphics[width=\linewidth]{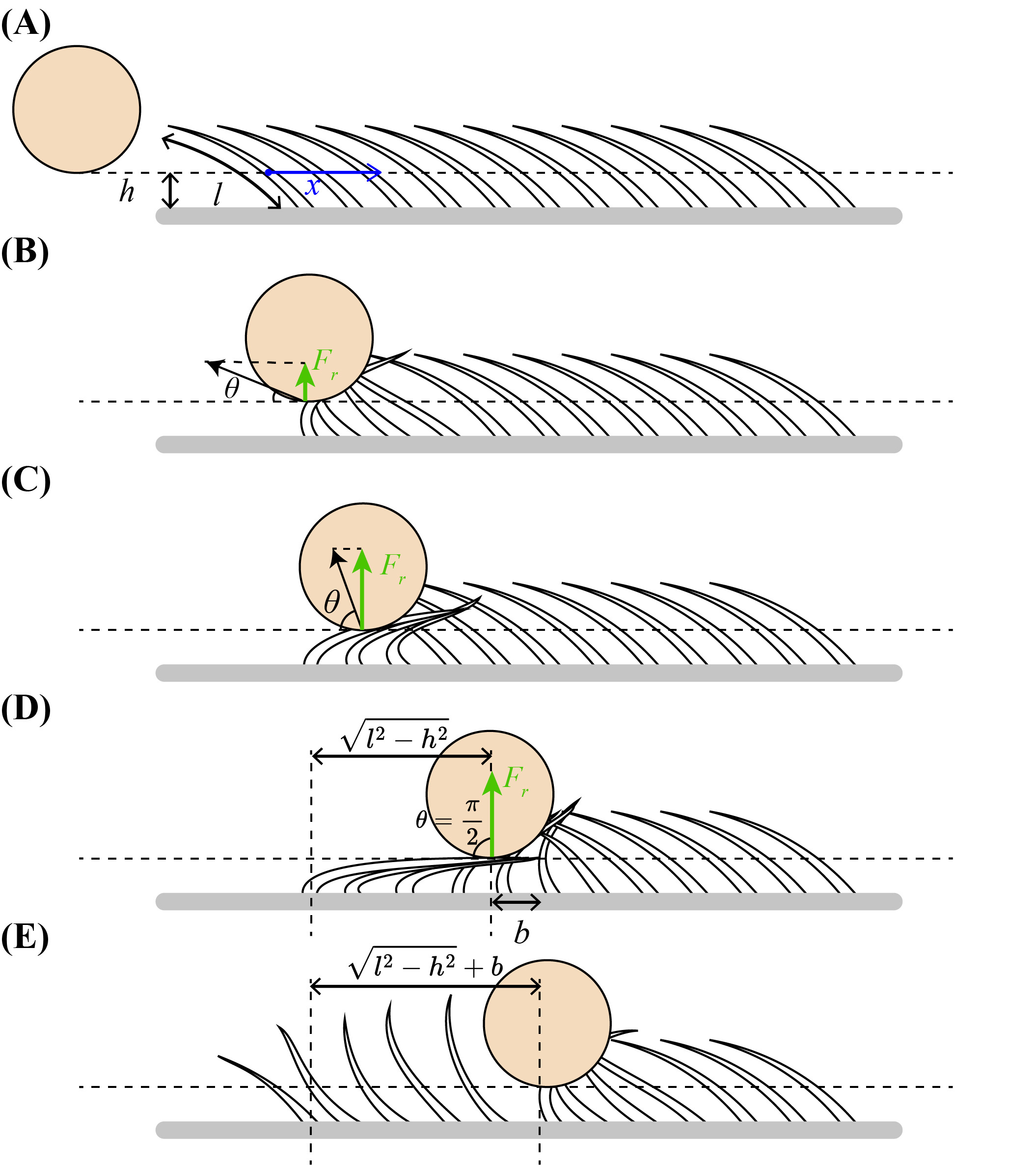}
    \caption{Schematic depicting fur behavior when stroked against the fur growth direction. (A) When the hand moves, (B) it first comes in contact with the initial fur bundle, (C) lifts it while entangling next bundles in the deformation phase, (D) slides across the bundle width during the rubbing phase, and (E) then releases it, showing a cyclic behavior.}
    \label{fig_haptic_model}
\end{figure}

First, we discuss the intensity design of the ultrasound haptic feedback based on the measurement results presented in \ref{subsec_measument_haptic}. 
When stroking the fur along the growth direction, the vertical reactive force $F_g(x)$ remained constant, as shown in Figure~\ref{fig_sens_result}, set at 0.6 for the intensity of ultrasound haptic feedback.
\begin{equation}
F_g(x) = F_0
\end{equation}
where $F_0$ is a constant.

In contrast, the reactive force exhibited a cyclic behavior when the fur was stroked against the growth direction.
The behavior is illustrated in Figure~\ref{fig_haptic_model}. 
The hand lifted, inverted, and released the hair bundles. Assuming hair length as $l$, width of the hair bundle as $b$, and hand height as $h$, 
the hair length when hair fell at the hand height was approximated by $\sqrt{l^2 - h^2}$.
The vertical reaction force $F_r$ during stroking in the reverse direction was considered a periodic function of $\sqrt{l^2 - h^2} + b$.

\begin{align}
    \label{eq_macromodel}
    F_r(x) &= F_r(x+\sqrt{l^2 - h^2} + b)
\end{align}

%\begin{CJK}{UTF8}{ipxm}周期の中(0<=x<=b+l)で，毛が接する角度θが0～π/2までxに対し線形に変化し，Π/2に到達したらその角度を維持したまま滑りを維持するという仮定のもと，\cite{Michel2015, Watzky2011}の一本の毛に対する式を，毛の束に拡張したものが以下である．kは比例定数である．\end{CJK}

Within period ($0 \leq  x \leq \sqrt{l^2 - h^2} + b$), the angle $\theta$ at which the hair came in contact with the skin changed linearly with $x$ from $0$ to $\frac{\pi}{2}$. Once it reached 
$\frac{\pi}{2}$, the angle was maintained during rubbing. Under this assumption, an existing formula for a single hair~\cite{Michel2015, Watzky2011} was extended to a bundle of hair, considering $k$ as a proportionality constant.

\begin{align}
    \label{eq_incycle}
    F_r(x) &= k\frac{\sin^2{\theta}}{(h\cos{\theta}+x\sin{\theta})^2} \ (0 \leqq x < \sqrt{l^2 - h^2}+b)\\\notag\\
    \theta &=   \begin{cases}
                    {\dfrac{\pi}{2}  \dfrac{x}{\sqrt{l^2 - h^2}}
                        \ (0 \leqq x < \sqrt{l^2 - h^2})}\notag\\\notag\\
                    {\dfrac{\pi}{2}
                        \ (\sqrt{l^2 - h^2} \leqq x < \sqrt{l^2 - h^2}+b)}\notag
                \end{cases}
\end{align}
    
%\begin{CJK}{UTF8}{ipxm}この式を図2の計測結果と同じ図にプロットしたものが図3である．計測結果と同様に，提案モデルは急な立ち上がりの後，緩やかな減少がサイクリックに起きているのを見ることができる．\end{CJK}

\begin{figure}[!t]
    \centering
    \includegraphics[width=\linewidth]{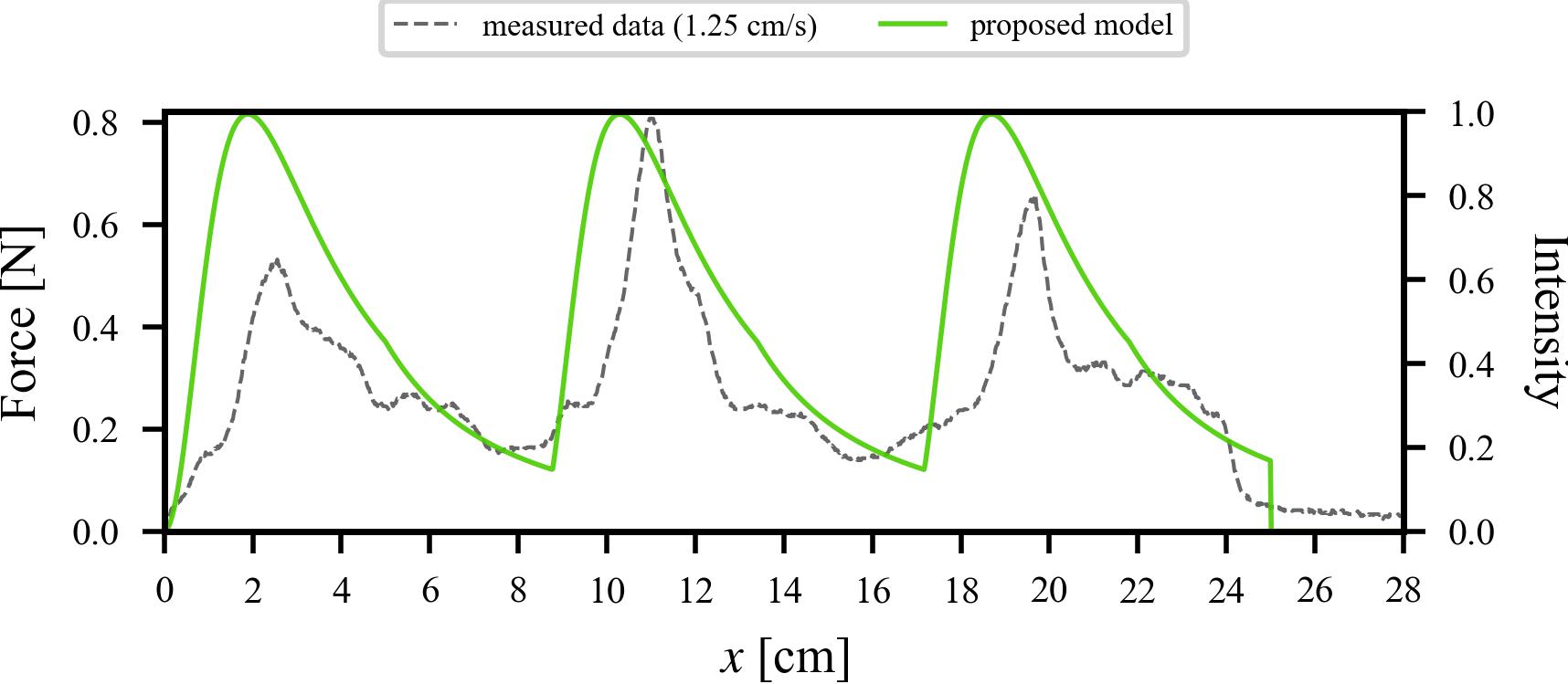}
    \caption{Comparison of the measured force at \SI{1.25}{cm/s} and the intensity of the speaker calculated by the proposed model when fur-stroking against the growth direction. The left y-axis represents force indicated by the black dashed line, while the right y-axis represents intensity of the ultrasound haptic feedback indicated by the green solid line. The x-axis shows the distance ($x$) from the initial point of contact. The proposed model demonstrates similar characteristics to the measured force, with a sharp rise and gradual decrease during the deformation and rubbing phases, respectively.}
    \label{fig_compare_model}
\end{figure}

Figure~\ref{fig_compare_model} shows the plots of the formulas along with the actual measurement results. The proposed model captured cycles that contained a pronounced initial increase and a subsequent gradual decrease in the deformation and rubbing phase, respectively, reflecting the measured behavior.

%\begin{CJK}{UTF8}{ipxm}次に周波数の設計について議論する．STMの円運動の周波数を変えることで，粗さ知覚が変わることが分かっている\cite{Ablart2019}．我々の知る限り．主観評価での粗さ知覚の変化のみ調べられている．円周20cmで30～80Haの範囲で粗さ知覚が変わることが分かっているため，順目では滑らかな触感提示のために80Hz，逆目では粗い触感提示のため30Hzを使う．\end{CJK}

Next, we discuss the frequency design. Existing studies have indicated that frequency alteration of the circular motion in STM influences roughness perception~\cite{Ablart2019}. To the best of our knowledge, previous studies have primarily focused on subjective evaluations of perception change.
Thus, roughness perception varies within a frequency range of approximately \SI{30}{Hz} to \SI{70}{Hz} for a circular path of 20 cm. Therefore, to simulate a smooth tactile sensation while stroking in the fur growth direction, we employed a frequency of \SI{70}{Hz}. Conversely, a frequency of \SI{30}{Hz} was used to evoke a rough tactile sensation when stroking against the fur growth direction.

\section{Experiment 1: Haptics}

%\red{\begin{CJK}{UTF8}{ipxm}[実験目的] 研究目的に対して，実験1は何を明らかにする立ち位置ですか？\end{CJK}}
This study aims to investigate the contribution of visuo-haptic stimuli based on an interactive model of hand movements to the experience of stroking voluminous fur in VR. 
When attempting to recreate the sensation of fur stroking, experimental biases including the visual aspects of fur and experimental intentions can influence the results.
Therefore, Experiment 1 was focused solely on haptic stimuli, where we examined the generation of fur sensation by our proposed haptic model of intensity and frequency providing further information.

In Experiment 2, we assessed the closeness with which the proposed visuo-haptic model replicated the experience of stroking fur compared to real fur by incorporating a visual CG model.

The Research Ethics Committee of the University of Tokyo approved the experiments presented in this and next sections (No. 23-559).

\subsection{Participants}

Eighteen participants (nine males and nine females) aged between 22 and 42 years, with an average age of 26.1, participated in this study. Among them, seventeen participants were right-handed, and three had prior exposure to ultrasound haptic feedback.

The participants were recruited under the pretext of examining mid-air ultrasound haptic feedback and tactile perception without any specific mention of fur. 
After the experiment, the participants were informed that the study focused on fur texture perception.

\subsection{Experimental Design}

An experiment was conducted to investigate the extent to which our proposed haptic model for intensity and frequency induced fur sensations without any prior expectations of fur texture. 
The experiment was framed as a survey of general tactile perception using mid-air ultrasound haptic feedback. 
The participants were debriefed after the experiment, clarifying that the study aimed to explore the sensations of fur.

\subsubsection{Experimental Conditions}
\begin{figure}[!htbp]
    \centering
    \includegraphics[width=\linewidth]{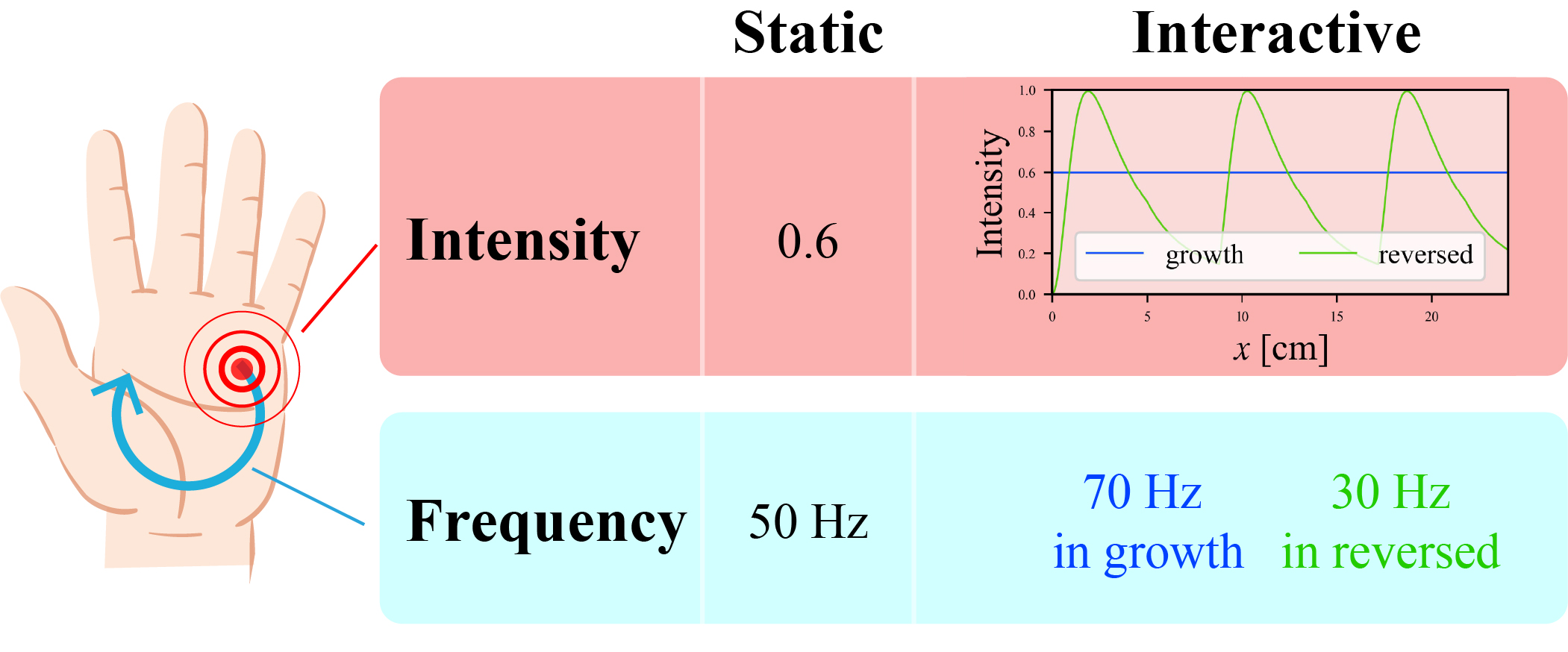}
    \caption{Summary of experimental conditions for two factors (intensity and frequency) and two levels (static and interactive).}
    \label{fig_ex01_conditions}
\end{figure}

The conditions of Experiment 1 were designed in combination with two levels for each factor of intensity and frequency of the mid-air ultrasound haptic feedback, resulting in a total of four conditions (Figrue~\ref{fig_ex01_conditions}). 
Both factors exhibited two levels: a static condition and an interactive condition.
Under the static condition for intensity factor, the intensity was fixed at 0.6, whereas the intensity under the interactive condition was set using the formula established in \ref{subsec_haptic_model}. 
Under the static frequency condition, the frequency was set at \SI{50}{Hz}. 
Under the interactive condition, the frequency was set at \SI{70}{Hz} in the growth direction and \SI{30}{Hz} in the reverse direction.
An intensity of 0.6 under the static condition was equivalent to the intensity in the growth direction under the interactive condition. 
The frequency of \SI{50}{Hz} set in the static condition for the frequency factor was an intermediate value used in previous studies on the roughness of ultrasound feedback~\cite{Wojna2023, Ablart2019} and has been utilized in several previous research~\cite{Freeman2021, Shen2023}. 
Within the frequency range of this experiment, no significant influence of frequency on the just noticeable difference (JND) of intensity was observed~\cite{Wojna2023}.

\subsubsection{Experimental Setup and System}

\begin{figure}[!htbp]
    \centering
    \includegraphics[width=\linewidth]{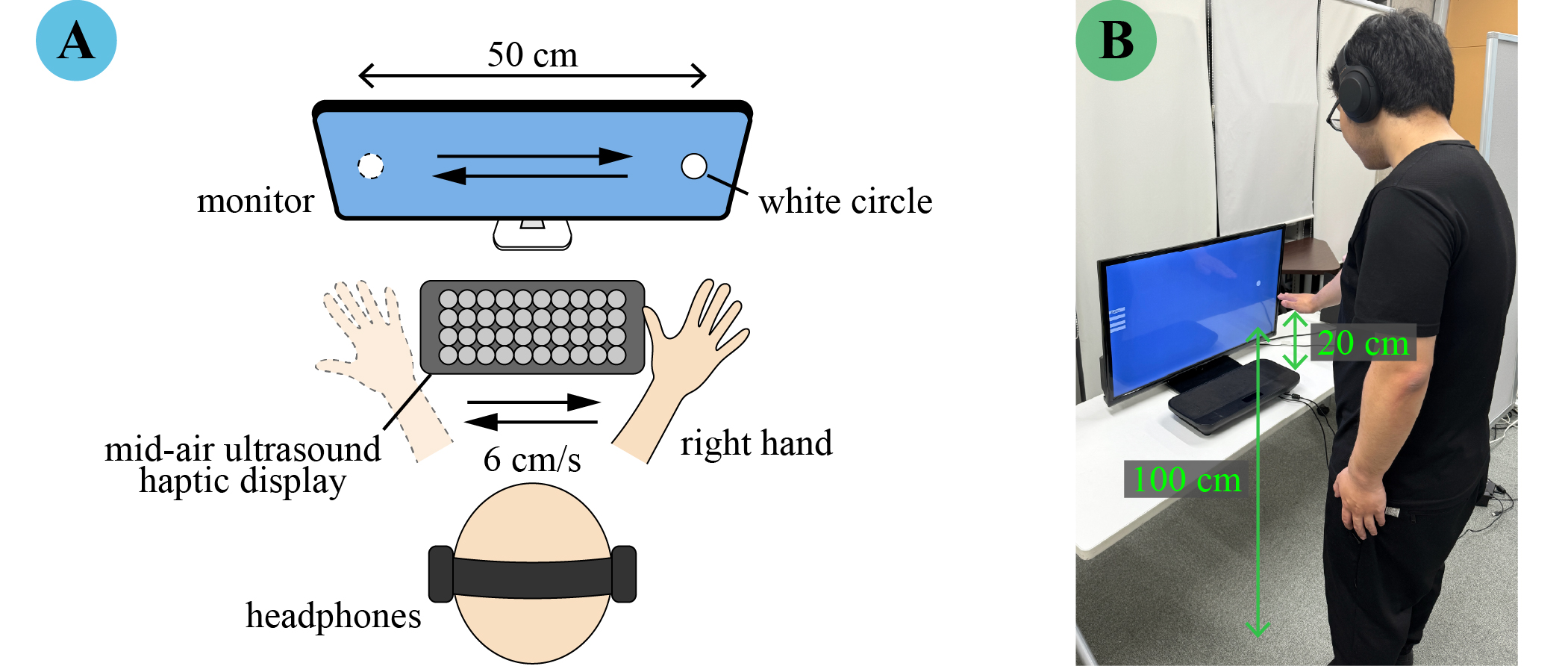}
    \caption{(A) Schematic representation of the setup for Experiment 1. (B) Environment of a participant engaged in Experiment 1. }
    \label{fig_setup_ex01}
\end{figure}

Figure~\ref{fig_setup_ex01} illustrates the setup and the system of the experiment.
For this experiment, a mid-air ultrasound haptic display, a monitor, a desk, and noise-canceling headphones (WH-1000XM4, SONY) were prepared. 

Haptic stimuli corresponding to the experimental condition were introduced as participants moved their right hand from right to left and from left to right. 
They were asked to choose one sensation that they felt was the closest among the twelve options.
Twelve response options were prepared based on previous studies and applications of texture presentation using mid-air ultrasound haptic feedback. 
The twelve options included 
``Cloth,’’ ``Wood,’’ ``Gel~\cite{Morisaki2021},’’ 
``Water,’’ ``Cork,’’ ``Metal~\cite{Roberto2023, Motoyama2022, Jang2020, Shen2022},’’ 
``Ice~\cite{Motoyama2022},’’ 
``Steam~\cite{Singhal2023, Wang2023},’’
``Skin~\cite{Marchal2020},’’
``Brick~\cite{Beattie2020},’’  ``Fur,’’ and ``Others (Free Answer).’’
This experimental concept was based on the experimental designs used in previous studies~\cite{Ho2007, Hosoi2023Wind}.

The hand movements of the participants were regulated by aligning them with moving markers displayed on a monitor. The method served two purposes: to ensure consistency in the stimuli presented among participants and to regulate biases in tactile perception during haptic exploration under different experimental conditions~\cite{Lederman1987}.
The right hand was positioned at a height of \SI{100}{cm} from the floor and \SI{20}{cm} from the mid-air ultrasound haptic display.  
The participants moved their right hand laterally over a distance of \SI{50}{cm} at a speed of \SI{6}{cm/s}. 
These values were determined by considering the operational range of the mid-air ultrasound haptic display and natural human movements without causing discomfort.

We conducted five trials for each condition, a total of 20 trials. 
The order of the 12 options was randomized for each trial, and the order of 20 trials was counterbalanced for each participant using a balanced Latin square~\cite{James1958}.

\subsection{Experimental Procedure}

The participants received a deceptive experimental explanation and answered a pre-experiment questionnaire. 
Subsequently, the participants wore noise-cancelation headphones and stood in a designated position in front of a mid-air ultrasound haptic display.
The experimental room was quiet. As white noise increases the perceived roughness of the ultrasound haptic feedback~\cite{Freeman2021}, no white noise was emitted from the headphones, activating only noise cancelation.

Each trial in Experiment 1 was conducted as follows. 
The participants extended their right hand horizontally, guided by the position of a white circle on the monitor. 
After a short pause, they moved their right hand \SI{50}{cm} from right to left at \SI{6}{cm/s} to match the movement of the moving white circle, stopping their hand at the position where the white circle stopped. 
After a 3-s pause, participants moved their right hands \SI{50}{cm} from left to right at \SI{6}{cm/s} to match the movement of the white circle, stopping their hand at the same position as the circle. 
Participants were instructed to focus on their right hand during this process.
After stopping their hands, participants answered a questionnaire on an iPad.

The procedure was repeated 20 times, with a 10-s break between trials. 
After completing all trials, the participants answered a post-experiment questionnaire. Thereafter, debriefing was conducted explaining that the main objective of the experiment was to feel the tactile sensation of fur. 
The duration of Experiment 1 was approximately 30 min.

\subsection{Result}

\begin{figure*}[!t]
    \centering
    \includegraphics[width=\linewidth]{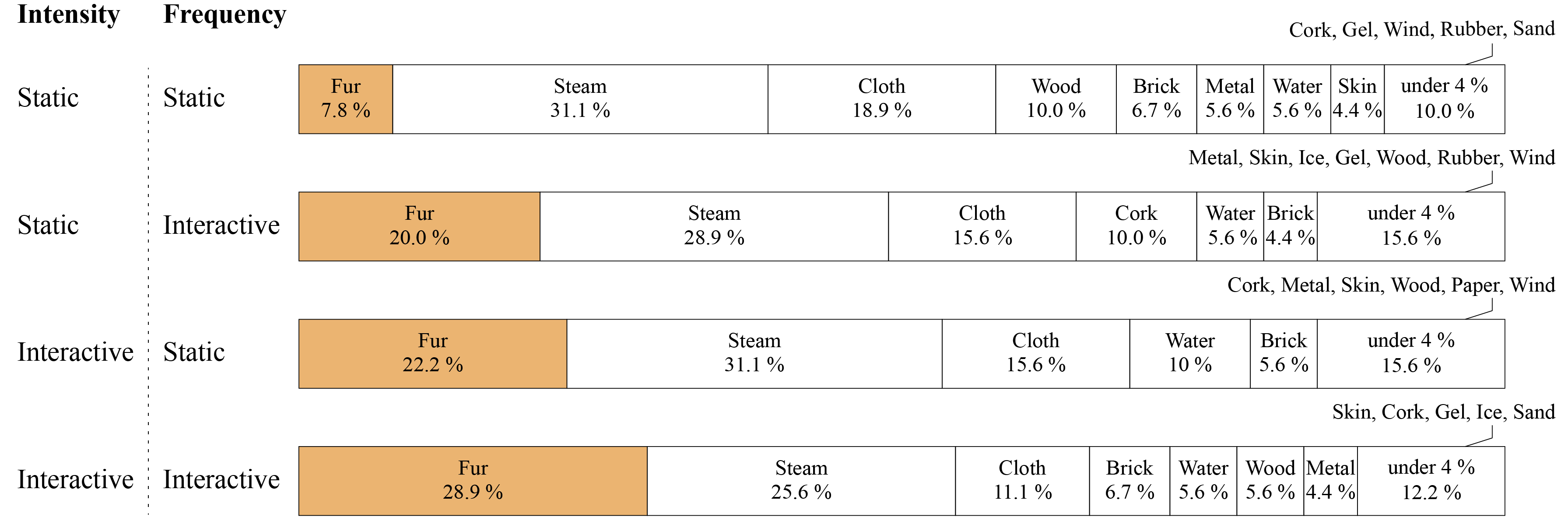}
    \caption{Results of the percentage of responses for the perceived closest material under four conditions of ultrasound haptic feedback. 
    ``Fur,’’ the main target of this research, is highlighted in brown, and those with a response rate of \SI{4}{\%} or less are summarized as ``under \SI{4}{\%}.’’}
    \label{fig_result_ex01}
\end{figure*}

Figure~\ref{fig_result_ex01} shows the percentage of responses for each experimental condition across all trials. The response rate for fur was positioned on the far left side, with the remaining options shown in descending order.
The results of statistical analyses are discussed further.
In this study, the critical p-value was set to 0.05 ($*: p<0.05, **: p<0.01$).

We first conducted the Shapiro--Wilk test to confirm data normality under each condition. Significant differences from the normal distribution were observed under all conditions.
Therefore, we performed an aligned rank transform (ART)\cite{Wobbrock2011} to enable us to conduct an ANOVA for non-parametric data. Hence, we conducted a two-way ANOVA. 
Table~\ref{table_result_ex01} summarizes the statistical test results.
The results exhibited significant main effects of intensity ($F (1, 51) = 10.4$, $p = 0.002$, ${\eta}^2 = 0.169$) and frequency ($F (1, 51) = 5.26$, $p = 0.026$, ${\eta}^2 = 0.094$). The interaction effect of intensity and frequency factors was not significant ($F (1, 51) = 0.284$, $p = 0.597$, ${\eta}^2 = 0.006$). 

\begin{table}[!htbp]
  \centering
  \caption{Results of the two-way ART ANOVA tests for response rate for fur. Bar graphs indicate the mean values of the conditions. $\ast : p < 0.05$, $\ast\ast : p < 0.01$.}
  \includegraphics[width=\linewidth]{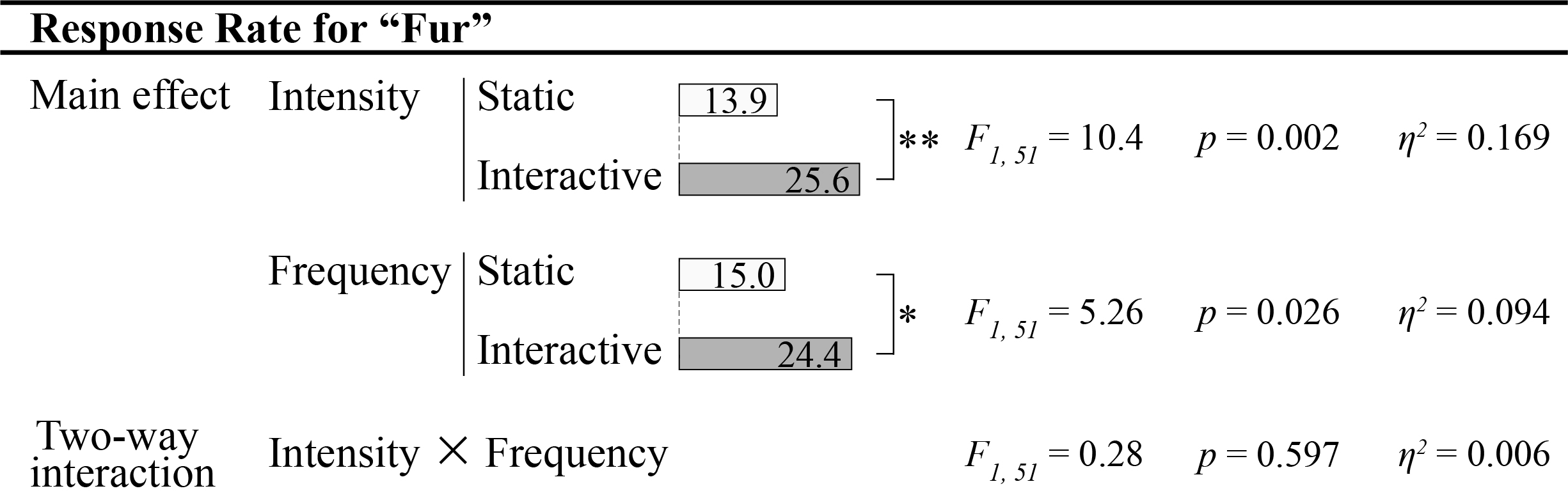}
  \label{table_result_ex01}
\end{table}

\subsection{Discussion}

First, we discuss the tactile sensation of ``fur’’, which was the main target of Experiment 1.
Statistical tests revealed significant main effects of intensity and frequency on the response rate for fur. 
It can thus be suggested that the proposed interactive design for the intensity and frequency of the ultrasound haptic feedback is effective in inducing the tactile sensation of fur. 
When both intensity and frequency of the ultrasound haptic feedback were set interactively, the highest response rate of \SI{28.9}{\%} was achieved under all experimental conditions; this value was also the largest among the response options for tactile sensations. 
Thus, ultrasound haptic feedback alone can induce a fur-like tactile sensation even without visual cues or prior contexts suggesting fur or animals.

Previous studies have explored various tactile sensations using ultrasound feedback. 
Although this study focuses mainly on the tactile sensation of fur, the results offer valuable insights into how ultrasound haptic stimulation is typically perceived. 
Across all the trials, the most frequently reported sensations were steam (\SI{29.2}{\%}), fur (\SI{19.7}{\%}), and cloth (\SI{15.3}{\%}). 
The trend is likely due to mid-air ultrasound haptic feedback, which is a gaseous medium, being better suited for simulating soft materials. 
Particularly, ``steam’’ was the only gas among the options and had the closest properties to the medium, which explains its highest response rate.
Post-experiment comments by participants included, ``I did not feel like I was touching a solid object’’ and ``I consistently felt a soft stimulus.’’ 
These findings provide valuable knowledge for future research and applications involving texture presentation using mid-air ultrasound haptic feedback.

% 条件間で統制していたが，手の動きによって絶対的な値は変わる可能性がある．

% 全体傾向の話，同じ参加者同じ条件下で選択肢がばらついている(ヒストグラム)．回答の確信度としては低い

\section{Experiment 2: visual and haptic}

For this study, we introduced a model for interactive visuo-haptic stimulation aimed at dynamically adjusting hand movements to simulate the tactile sensation of stroking voluminous fur. Experiment 2 was performed to assess the impact of interactive visuo-haptic models on the fur-stroking experience. 
In the experiment, we compared our proposed model with real fur to assess the realism of it simulating the experience of stroking voluminous fur.

\subsection{Participants}

Twenty-one participants (13 males and 8 females) aged between 22 and 47 years, with an average age of 26.9, participated in the experiment. Among them, 19 were right-handed. Three had prior exposure to ultrasonic vibrators and 18 had prior VR experience.
The sample size was calculated to satisfy a significance level of 0.05, effect size of 0.5, and power of 0.6.

\subsection{Experimental Design}

The experimental conditions for Experiment 2 were a combination of two levels for each factor of the visual CG model (\textit{Visual}) and the haptic model (\textit{Haptic}) of the mid-air ultrasound haptic feedback, resulting in a total of four conditions. 
Both factors exhibited two levels: a static model (\textit{static condition}) and an interactive model (\textit{interactive condition}). 
% 視覚要因の静的な条件では，毛並みのCGモデルは手との接触によって変形せず，動的な条件では毛並みのCGモデルは3.2節で述べたように手によって変形を起こす．触覚要因の静的な条件では，超音波触覚刺激の強度と周波数は手の移動によって変化させず，実験1のA条件と同様の刺激を提示した．一方で触覚要因の動的な条件では，超音波触覚刺激の強度と周波数は手の移動によって変化さえ，実験1のB条件と同様の刺激を提示した．

Under the static condition of the visual factor, the fur CG model remained undeformed on contact with the hand, while under the dynamic condition, as described in Section 3.2, the fur CG model underwent deformation by the hand.
Under the static condition of the haptic factor, the intensity and frequency of the ultrasound haptic feedback remained static with hand movement, consistent with the \textit{Intensity-Static \& Frequency-Static} condition in Experiment 1. Under the dynamic condition of the haptic factor, the intensity and frequency of the ultrasound haptic feedback varied with the hand movement based on the proposed haptic model, mirroring the \textit{Intensity-Interactive \& Frequency-Interactive} condition in Experiment 1.

To establish a basis for assessing the tactile experience of stroking fur, participants were instructed to stroke real fur horizontally before receiving visuo-tactile stimulation in each trial. 
The real fur used was the same as that used in \ref{subsec_measurment}, and the position of the fur was set at the same height as the position for presenting the ultrasound haptic feedback to align with the posture and hand movements during the stimuli.
Following real fur stroking, the participants were exposed to visuo-tactile stimuli based on the experimental conditions. Subsequently, they completed a questionnaire assessing their perception of the presented stimuli. Each evaluation item was scored on a 10-point scale and encompassed the reality of the fur-stroking experience (1: does not feel like stroking fur at all; 10: fully feels like stroking fur), perceived softness (1: not soft at all; 10: fully soft), comfort of the fur (1: not comfortable at all; 10: fully comfortable), and enjoyment during the trial (1: did not enjoy at all; 10: fully enjoyed).
The questionnaire was completed using an iPad.

We conducted one trial for each condition, with a total of four trials.
The order of the four trials was counterbalanced for each participant using a balanced Latin square~\cite{James1958}.

\subsection{Procedure}

\begin{figure}[!htbp]
    \centering
    \includegraphics[width=\linewidth]{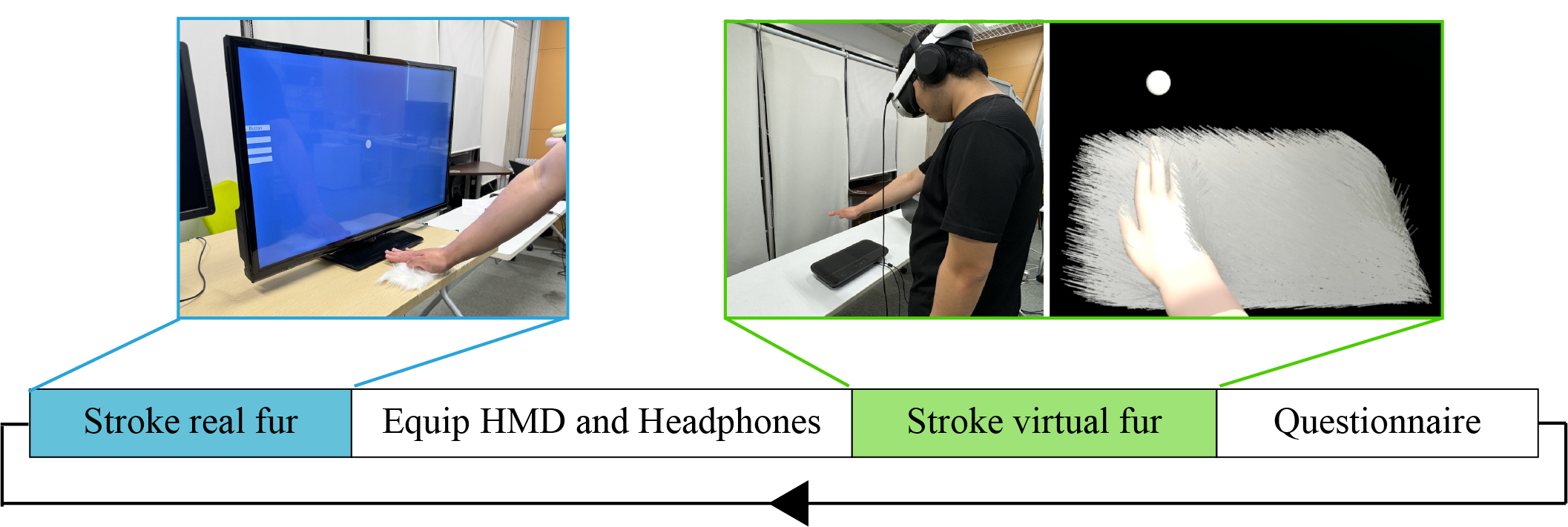}
    \caption{Flow of each trial in Experiment 2.}
    \label{fig_ex02_procedure}
\end{figure}

The participants received an explanation of the experiment and they provided written informed consent. Each trial was performed in accordance with the experimental conditions. 
The flow of each trial is illustrated in Figure~\ref{fig_ex02_procedure}. 

The contents of each trial included: first, the participants stood in front of a real fur placed on a 100 cm-high stand. 
Then, they stroked the fur from left to right and from right to left, aligning it with a moving white circle on the monitor, replicating Experiment 1. 
Following the stroke, they stood in front of the ultrasound transducer array wearing an HMD (Meta Quest 3, Meta) and noise-cancelation headphones (WH-1000XM4, SONY). 
The participants placed their hands to the right of the fur using the position of a white guide ball on the VR space and moved their hands from right to left in the natural direction of fur growth, synchronizing with the ball movement. 
After reaching the left end, they stopped for 3 s and then moved their hands from left to right in the reverse direction, mirroring the motion of stroking fur. 
Subsequently, they removed the HMD and headphones. Then, they completed a questionnaire on an iPad. A 1-min break was given between each trial. 

After completing all trials, the participants completed a post-experiment questionnaire. The duration of Experiment 2 was 30 min.

\subsection{Result}

\begin{figure*}[!t]
    \centering
    \includegraphics[width=\linewidth]{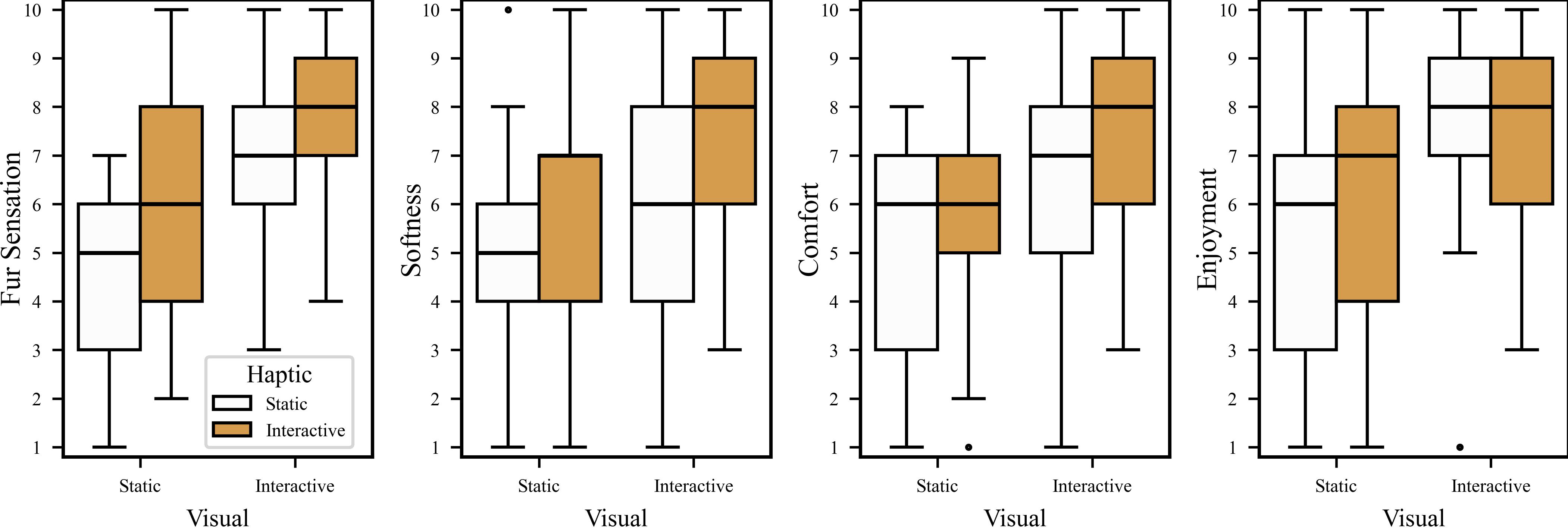}
    \caption{Box plot of ``Fur Sensation,’’ ``Softness,’’ ``Comfort,’’ and ``Enjoyment.’’ Data points that are away from the median by more than 1.5 times the interquartile range are denoted as outliers.}
    \label{fig_result_ex02}
\end{figure*}

Figure~\ref{fig_result_ex02} shows a box plot of the 10-point scale scores for fur sensations, softness, comfort, and enjoyment.
We performed ART~\cite{Wobbrock2011} and conducted a two-way ANOVA. Table~\ref{table_result_ex02} summarizes the statistical test results. 

\subsubsection{Fur Sensation}
For fur sensation, the results showed a significant main effect of visual factor ($F(1, 60)=58.1$, $p < 0.001$, ${\eta}^2 = 0.492$) and significant main effect of haptic factor ($F(1, 60) = 12.6$, $p < 0.001$, ${\eta}^2 = 0.174$). 
The interaction effect of visual and haptic factors was not significant ($F (1, 60) = 1.06$, $p = 0.307$, ${\eta}^2 = 0.017$).

\subsubsection{Softness}
Regarding the perception of softness, the results showed a significant main effect of visual factors on fur sensation ($F(1, 60) = 13.3$, $p < 0.001$, ${\eta}^2 = 0.181$) and a significant main effect of haptic factor ($F (1, 60) = 5.87$, $p = 0.018$, ${\eta}^2 = 0.089$). 
The interaction effect of visual and haptic factors was not significant ($F (1, 60) = 0.209$, $p = 0.649$, ${\eta}^2 = 0.003$).

\subsubsection{Comfort}
For the sensation of comfort, the results showed a significant main effect of visual factors on fur sensation ($F (1, 60) = 18.2$, $p < 0.001$, ${\eta}^2 = 0.233$) and no significant main effect of haptic factor ($F (1, 60) = 1.50$, $p = 0.226$, ${\eta}^2 = 0.024$). 
The interaction effect of visual and haptic factors was not significant ($F (1, 60) = 0.180$, $p = 0.673$, ${\eta}^2 = 0.003$).

Previous studies have suggested an association between softness and comfort~\cite{Pasqualotto2020, Kitada2021, Xue2023}.
Therefore, to assess the relationship between the perception of softness and comfort, Spearman's rank-order correlation was conducted. The analysis revealed a strong positive correlation between the two variables, $r_s = 0.86$, which was statistically significant ($p < 0.001$).

\subsubsection{Enjoyment}
For the sensation of enjoyment, the results showed a significant main effect of visual factors on fur sensation ($F (1, 60) = 15.5$, $p < 0.001$, ${\eta}^2 = 0.206$) and no significant main effect of haptic factor ($F (1, 60) = 1.91$, $p = 0.172$, ${\eta}^2 = 0.031$). 
The interaction effect of visual and haptic factors was not significant ($F (1, 60) = 0.382$, $p = 0.539$, ${\eta}^2 = 0.006$).

\begin{table}[tbp]
  \caption{Results of the two-way ART and ANOVA tests for ``Fur sensation,’’ ``Softness,’’ ``Comfort,’’ and ``Enjoyment.’’ Bar graphs indicate the median values of the conditions $\ast : p < 0.05$, $\ast\ast : p < 0.01$.}
  \centering
  \includegraphics[width=\linewidth]{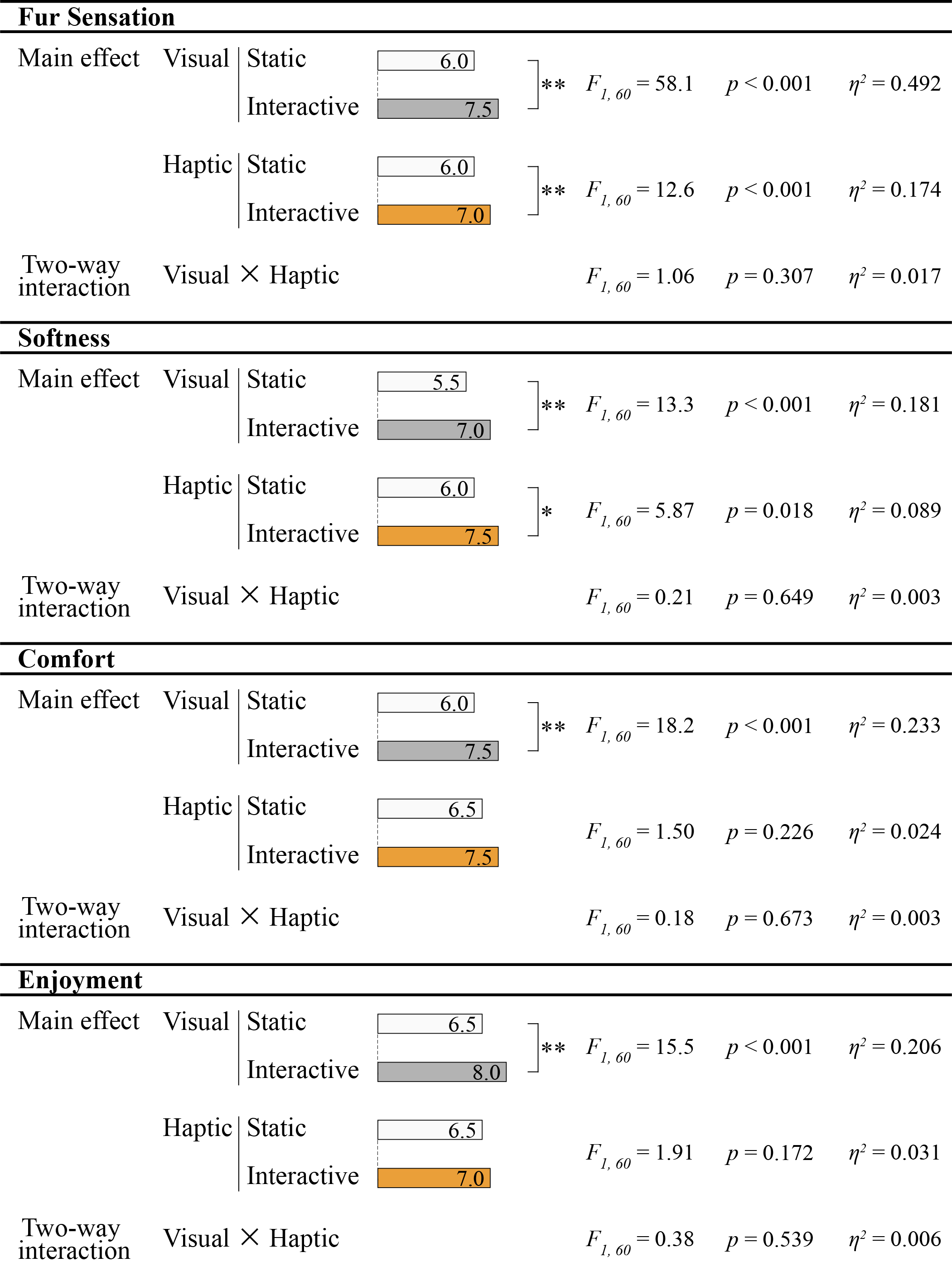}
  \label{table_result_ex02}
\end{table}

\subsection{Discussion}

% 毛並み触感のリアリティについて，視覚要因と触覚要因のそれぞれについて有意な主効果が見られた．これは実際の毛並みと比較した場合においても，視覚・触覚の両方について提案したインタラクティブデザインが毛並み体験のリアリティに対して有効であることを意味する．触覚刺激の有意な主効果は、撫でる方向に依るボリュームのある毛並みにおける毛の曲がり方の違いとその曲がり方による反力と粗さの違いを提示することが、毛のボリューム感の提示を達成したと解釈できる。この知見は、後に述べる柔らかさ知覚の向上からも支持される。どちらもインタラクティブな場合がもっとも高い中央値，平均値をとり，その条件での体験について事後アンケートでは「映像と刺激が一致することで毛並みをより感じやすくなったように感じた。」「VRと毛の接触感覚は非常に人間の感覚と近いものがあった。」といった回答が見られた． 
Considering the fur sensation, significant main effects were observed for both visual and haptic factors. 
Thus, the proposed interactive design is effective in enhancing the reality of voluminous fur experience in terms of both visual and haptic aspects, even when compared to actual voluminous fur.
The significant enhancement of the fur sensation by interactive haptic design can be interpreted as successfully achieving the presentation of fur volume by displaying differences in the bending of the fur and the corresponding variations in reactive force and roughness based on the direction of stroking.
These findings were also supported by the enhancement in the perception of softness, which will be discussed later.
The highest median at 8.0 and mean values at 7.8 were obtained when both factors interacted. In the post-experiment questionnaire, participants reported this condition with comments, ``The alignment of visual and haptic stimuli made the fur sensation more pronounced’’ and ``The tactile sensation of fur in VR was very close to that of real fur.’’

% 一方で，実際の毛皮と比べての相違点について，主に2点の観点での自由回答があった．一つ目は毛皮のベースである皮の部分についてである．本研究では，毛皮の毛の部分についての提示のみを行っており，毛が生えている皮部分のしっかりとした抵抗は超音波振動フィードバックでは提示できていない．「VR空間ではダチョウの羽毛を集めた塊に手を通しているような感覚だった．」「映像と感触が両方あわさったらかなり毛皮っぽくなってびっくりしました．全体的にふわふわ感は感じ，サラッサラの毛の先っぽの方「だけ」をすーっと撫でてる感覚で，ただその奥の毛が生えてる奥にある実体のある感じがあるともうちょっとリアルなのかなと思いました．」．2つ目は温度感についてである．先行研究では超音波刺激とミストを用いて冷覚刺激を提示する手法があるが，本研究では触覚刺激のみを提示しており，実際の毛のひんやり感は提示できていない．
However, two main points of free responses regarding the differences compared to actual fur were reported. The first concerned the base of the fur or the skin. In this study, only fur was presented and the resistance provided by the underlying skin could not be presented using mid-air ultrasound haptic feedback. 
This limitation was noted in comments, such as ``In VR, it felt like passing my hand through a bundle of ostrich feathers’’ and ``Combining visual and tactile sensations made it feel surprisingly fur-like. The overall fluffy sensation and smooth tips were noticeable, but the absence of a solid feeling in the underlying skin made it less realistic.’’
The second point concerned thermal sensation. Previous studies have used ultrasonic stimulation and mist to present cold sensations~\cite{Nakajima2020, Nakajima2021, Motoyama2022}; however, our study only presented tactile feedback and we did not present the actual coolness of fur.

% 毛並みの柔らかさ知覚について，視覚と触覚のどちらの要因にも有意な主効果が見られた．これは，視覚的変形と触覚的な力の変化のそれぞれが毛並みの柔らかさを向上させることを意味する．柔らかさ知覚に関する先行研究において，視覚・触覚それぞれが寄与することが分かっており，本実験結果もこの知見に一致している．
% 心地よさは視覚のみの有意な主効果がみられ，触覚刺激について有意な主効果は見られなかった．また追加の統計検定の結果，柔らかさの知覚と心地よさの分布には強い正の相関があった．これは，柔らかさと心地よさ･快感に関する先行研究の知見とも合致する．触覚刺激のインタラクティブデザインは柔らかさの観点では心地よさに寄与したと考えられるが，超音波刺激の気泡によるぷつぷつ感が心地よさに対してネガティブな影響を与えたため，心地よさに対して統計的な有意差が見られなかったと考えられる．
Regarding the perception of fur softness, significant effects were observed for both visual and haptic factors. Thus, visual deformation and haptic changes both enhance the perception of fur softness. 
The obtained results for softness agreed with the findings of previous studies, where displacement, force distribution, and contact area when pressing an object influenced the perception of softness~\cite{Drewing2009, Cellini2013, Tiest2014, Punpongsanon2015}. 
Focusing on haptic modality, our study dynamically changed the reactive force in response to hand movements without changing the spatial distribution of haptic stimuli. Increasing the intensity of the ultrasound haptic feedback in response to displacement during reversed stroking can contribute to softness perception, similar to that of a spring model~\cite{Marchal2020}.

Significant main effects on comfort and enjoyment were observed for visual factors but not for haptic factors. Further statistical tests revealed a strong positive correlation between the perceptions of softness and comfort. The finding was consistent with those of previous studies on the relationship between softness and comfort~\cite{Pasqualotto2020, Xue2023}. 
While the interactive design of haptic stimuli contributed to comfort from the perspective of softness, the prickly sensation caused by air bubbles in the ultrasound feedback likely had a negative impact on comfort and enjoyment during fur interaction.
In the post-experiment questionnaire, a comment read, ``Ultrasound stimulation was more tingling than the actual tactile sensation of the fur.’’
Recent studies have indicated that multi-point STM can deliver a mild tactile sensation suitable for relaxation~\cite{Shen2023}. For applications focusing on comfort, the proposed model requires an adjustment in the number of focal points.

% 心地よさにはコンテキストが大事

\section{Limitation and Future Work}

%\begin{CJK}{UTF8}{ipxm}手の動作の制約・毛の種類・提示範囲形状\end{CJK}

% 本研究はボリューミーな毛並みの疑似的な提示を検証した初の研究であり，毛並み触感の生起そのものを研究スコープとしている．そのため，毛並み触感提示についていくつかのリミテーションが存在する．まず，一つ目として手の動きについてである．本研究では，毛並みとのインタラクションにおいて最も一般的であると考えられる，左右方向の撫で動作に手の動きを限定している。一方で，ボリューミーな毛並みとの触覚的なインタラクションには，指でつまむ動作や垂直方向に押し込む動作など多岐にわたる．手の動きと合わせて考えるべきリミテーション2つ目は，触覚刺激の提示部位・位置についてである．本研究では超音波刺激提示として基本的とされる掌への円状のSTMを用いたが，毛は一本一本が細かいため指の間や手の側面などに対しても触覚刺激が提示される．3つ目は毛の種類についてである．本研究では，測定のサンプルとして用いたリアルファーを元に毛並みモデルのパラメータを設定し実験を行った．以上に挙げた3つのリミテーションに対し，モデルをアップデートすることで，Future Worksでは，よりリアルで多様な毛並みインタラクションの体験を実現できると考えられる．

This study is the first to investigate the simulated presentation of voluminous fur by focusing on the induction of fur sensation. 
Our study had several limitations. 
First, regarding hand movements, our study limited hand interactions to stroking, which is considered the most common interaction with fur. 
However, interactions with voluminous fur can involve various hand movements, such as pinching or pressing vertically. The position of the haptic stimulation is also a limitation. Our study used a circular STM on the palm, which is a basic method for presenting ultrasound haptic feedback. 
However, fur comprises fine hair strands that interact with the spaces between the fingers or sides of the hand. 
Third, concerning the type of fur used, our experiments set the parameters for visual and haptic modalities based on a simple voluminous fur sample. 
Although our method is effective for soft fur, presenting the tactile sensation of brush-like stiff hair is difficult owing to the output limitation of ultrasonic transducers. 
By adjusting the parameters of the proposed model, we can achieve the experience of stroking soft fur that covers the entire palm. However, other types of fur interactions require modifications to the model.

Updating the model to address these limitations can enable more realistic and diverse fur interactions in future studies. 
Furthermore, verifying the effectiveness of presenting various tactile fur sensations in animal-assisted therapy, entertainment, and online shopping applications is crucial.

\section{Conclusion}

%\red{\begin{CJK}{UTF8}{ipxm}なにしたんですか？\end{CJK}}

%\red{\begin{CJK}{UTF8}{ipxm}RQへの回答\end{CJK}}

%\red{\begin{CJK}{UTF8}{ipxm}予想されるインパクト\end{CJK}}

This paper presents a novel approach for simulating the experience of stroking voluminous fur using an interactive visuo-haptic model in virtual reality.
Based on the observations and measurements of stroking fur, we focused on the anisotropy of voluminous fur and developed interactive models for both visual and haptic modalities.
By employing a mid-air ultrasound haptic feedback system and detailed CG models of fur, we achieved a realistic replication of the soft, voluminous texture of fur. 
Our experiments validate the effectiveness of the proposed model in evoking the sensation of voluminous fur stroking, with significant contributions from the interactive models of both visual and haptic feedback. 
Despite limitations such as the absence of underlying skin resistance and thermal sensations, the results indicate promising applications for enhancing user experiences in VR environments. Future studies will focus on addressing existing limitations and exploring additional applications in therapeutic and entertainment domains.

\
\bibliographystyle{IEEEtran}
\bibliography{ref}

\begin{IEEEbiography}[{\includegraphics[width=1in,height=1.25in,clip,keepaspectratio]{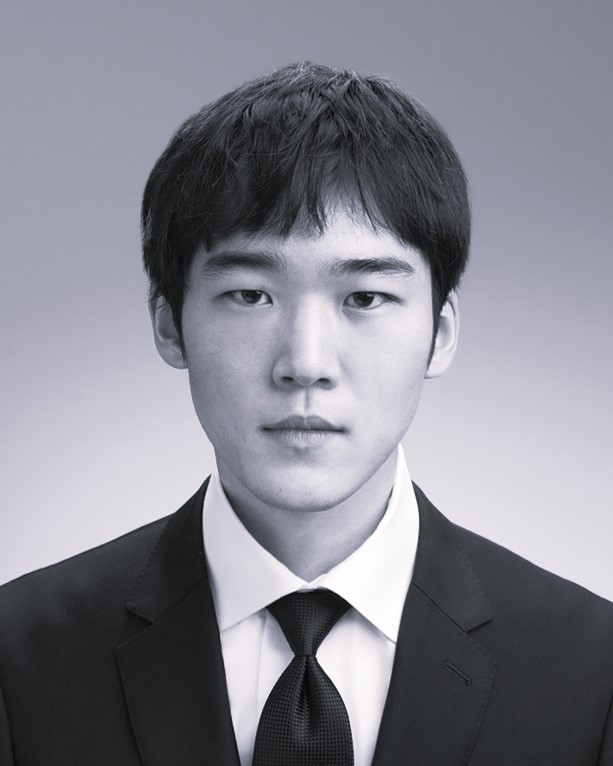}}]{Juro Hosoi} received the Master of Environmental Studies degree from the Graduate School of Frontier Sciences, the University of Tokyo, Tokyo, Japan in 2023.
He is a doctoral student of the Graduate School of Frontier Sciences, the University of Tokyo now.
He is also awarded the JSPS Research Fellowship for Young Scientists (DC1) from the Japan Society for the Promotion of Science from 2023 to 2026.
His current research focuses on reproducing various tactile sensations.\end{IEEEbiography}

\begin{IEEEbiography}[{\includegraphics[width=1in,height=1.25in,clip,keepaspectratio]{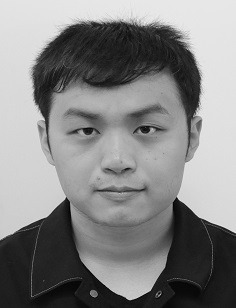}}]{Du Jin} received his bachelor of engineering degree from the Faculty of Automation, Harbin Engineering University in 2020. He is a master student of the Graduate School of Frontier Sciences, the University of Tokyo now. His research interest is user interface design in virtual reality.\end{IEEEbiography}

\begin{IEEEbiography}[{\includegraphics[width=1in,height=1.25in,clip,keepaspectratio]{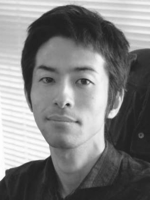}}]{Yuki Ban} received the M.S. and Ph.D. degrees in information science and technology from The University of Tokyo, Japan, in 2013 and 2016, respectively. From 2016 to 2017, he was a Researcher with Xcoo Inc. Research. Since 2017, he has been with The University of Tokyo. He is currently a Project Lecturer with the Department of Frontier Sciences, The University of Tokyo. His current research interests include cross-modal interfaces and biological measurement.
\end{IEEEbiography}

\begin{IEEEbiography}[{\includegraphics[width=1in,height=1.25in,clip,keepaspectratio]{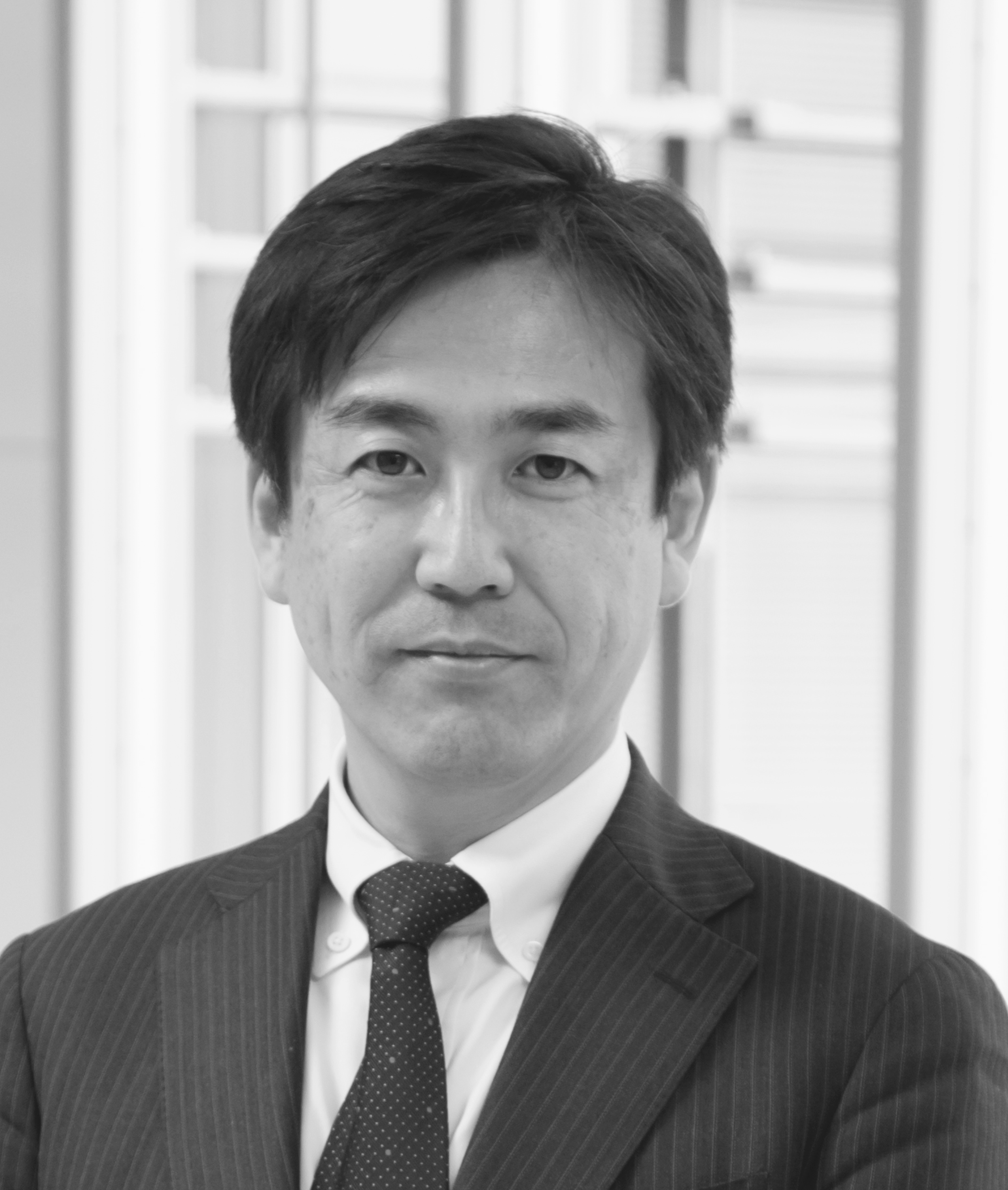}}]{Shin'ichi Warisawa} is a professor 
at the Department of Frontier Sciences at the Univ. of Tokyo. 
From 1994 to 2000, he served as an Assistant Prof. of Tokyo Institute of Technology, Since 2000 he has been working at the University of Tokyo. 
He was a visiting researcher at Massachusetts Institute of Technology from 2010 to 2011, and a visiting professor at Université Jean Monnet in 2016. 
His current research focuses on wearable/ambient human health monitoring. 
Research cores are nano/micro sensing devices fabrication and sensing information technology application for human well-being.\end{IEEEbiography}

\vfill

\end{document}